    \documentclass[final,3p,times,twocolumn,sort&compress]{elsarticle}
\usepackage{amssymb,amsmath}
\usepackage{lineno,hyperref}
\usepackage{arydshln}
\modulolinenumbers[5]

\journal{Journal of Molecular Liquids}

\newcommand{\beq}{\begin{equation}}
\newcommand{\eeq}{\end{equation}}

\newcommand\beqa{\begin{eqnarray}}
\newcommand\eeqa{\end{eqnarray}}
\newcommand{\nn}{\nonumber\\}
\newcommand{\la}{\lambda}
\newcommand{\e}{\eta}

\def\bal#1\eal{\begin{align}#1\end{align}}
\newcommand{\XX}{X}

\usepackage[usenames]{color}

\newcommand{\kk}{0.15\textwidth}

\newcommand{\hh}{36mm}

\newcommand{\hhb}{39mm}

\begin{document}

\begin{frontmatter}



\title{Structural and thermodynamic properties of fluids whose molecules interact via one-, two-, and three-step potentials}


%\author{}

%\affiliation{organization={},%Department and Organization
            %addressline={},
            %city={},
            %postcode={},
            %state={},
            %country={}}

\author[1]{Santos B. Yuste}
\ead{santos@unex.es}
\author[1]{Andr\'es Santos\corref{cor1}}
\ead{andres@unex.es}
\author[2]{Mariano L\'opez de Haro}
\ead{malopez@unam.mx}

\cortext[cor1]{Corresponding author}

\address[1]{Departamento de F\'{\i}sica, Universidad de
Extremadura and Instituto de Computaci\'on Cient\'ifica Avanzada (ICCAEx), E-06006 Badajoz, Spain}
\address[2]{Instituto de Energ\'ias Renovables, Universidad Nacional Aut\'onoma de M\'exico
(U.N.A.M.), Temixco, Morelos 62580, M{e}xico}

\begin{abstract}
The structural and thermodynamic properties of fluids whose molecules interact via potentials with a hard-core plus a square well, a square shoulder, and a second square well, are considered. Those properties are derived by using a  (semi-analytical) rational-function approximation method as a particular case of the more general formulation provided earlier involving potentials with a hard-core plus $n$ piecewise constant sections.
Comparison of the  results with recent simulation data confirms the usefulness of the approach.
\end{abstract}




\end{frontmatter}


\section{Introduction}
\label{sec1}

Undoubtedly, the late Douglas Henderson was a recognized international leader in the theory of liquids. In this area he provided significant contributions, such as the one with John Barker concerning the perturbation theory of fluids. In particular, in their outstanding paper {\it What is ``liquid''? Understanding the states of matter}, published in Reviews of Modern Physics \cite{BH76}, Barker and Henderson clearly laid out the ground for further developments in liquid-state theory. There they wrote {\it ``The aim of the physics of liquids is to understand why particular phases are stable in particular ranges of temperature and density (phase diagrams; \ldots), and to relate the stability, structure, and dynamical properties of fluid phases to the size and shape of molecules, atoms, or ions and the nature of the forces between them (which in turn are determined by the electronic properties).''} In this regard, the discrete potentials including the square-well (SW), the triangle-well, or the square-shoulder (SS) potentials, as well as potentials with a hard-core plus piecewise constant regions, have been important in advancing in this direction as simple models of intermolecular interaction. Using such potentials, problems such as liquid-liquid transitions \cite{FMSBS01,SBFMS04,MFSBS05,CBR07}, colloidal interactions \cite{GCC07}, the anomalous density behavior of water and supercooled fluids \cite{OFNB08,ONB09}, and the thermodynamic and transport properties of Lennard-Jones fluids \cite{CMA84,CSD89}, among others, have been successfully addressed.

Very recently Perdomo-P\'erez {et al.} \cite{PMPSGLVTC22} have considered fluids with competing interactions modeled through a succession of SW and SS potentials. Their main goal was to systematically study the effect of each potential contribution on the physical properties of a competing interaction fluid. Since the discrete potentials are discontinuous and they cannot be directly used in simulation methods that require the explicit calculation of the force at the breakpoints,  they further determined suitable continuous and differentiable potentials that, apart from mimicking some of the properties of their discrete potential counterparts, had the same reduced second virial coefficient. With this approach, they have produced new simulation data on both the radial distribution function (RDF) and the compressibility factor of fluids interacting via those discrete-like potentials,  thus extending and complementing the (relatively scarce) earlier simulation data \cite{GCC07,HTS09,HTS11,BOO11} on these properties.

In previous work \cite{SYH12}, we used the rational-function approximation (RFA) methodology \cite{S16}, which is a semi-analytical approach, to derive the general formulae for the structural properties of fluids whose
molecules interact via discrete potentials with a hard core plus
an arbitrary number of piecewise constant sections of different
widths and heights, and compared the outcome with the then available computer simulation results \cite{BOO11}. This work led later to the consideration of the special case in which the fluids were constituted by molecules whose interaction potential consisted of a discrete potential with a hard core plus different combinations of a repulsive shoulder and an attractive well \cite{SYHBO13}. The question then naturally arises as to whether our theoretical development will also perform well when compared to the new simulation data involving fluids interacting via a potential with a hard-core plus a SW, a SS, and a second SW. The aim of this paper is precisely to evaluate such performance.

The paper is organized as follows. In order to make it self-contained, in Section \ref{sec2} we provide the background material of the RFA approach for discrete fluids whose intermolecular potential has a hard-core plus $n$ piecewise constant sections. This is followed by Section \ref{results}, which contains the results of our calculations for illustrative cases and their comparison with the recent simulation data  for such cases \cite{PMPSGLVTC22}. We close the paper in Section \ref{concl} with further discussion and some concluding remarks.

\section{The rational-function approximation}
\label{sec2}
In this section we provide a brief but self-contained account of the RFA methodology as applied to our fluid whose molecules interact through a potential with a hard-core plus $n$ piecewise constant regions. For a more detailed account, we refer the reader to Refs.~\cite{SYH12,S16}.

\subsection{Formal expressions}

Let us recall first some general relations between the thermodynamic and structural properties of fluids. The virial route to the equation of state  leads to  \cite{BH76,HM13,S16}
\bal
\label{presMul}
Z\equiv \frac{p}{\rho k_B T}
=&1-\frac{\rho}{6k_BT}\int d \mathbf{r}\, g(r) r \frac{\partial \phi(r)}{\partial r}\nn
=&1+\frac{\rho}{6}\int d \mathbf{r}\, y(r) r \frac{\partial e^{-\phi(r)/k_BT}}{\partial r},
\eal
where $Z$ is the compressibility factor, $p$ is the pressure, $\rho$ is the number density, $k_B$ is the Boltzmann constant,  $T$ is the absolute temperature,  $\phi(r)$ is the interaction potential (assumed to be  spherically symmetric and  pairwise additive),  $g(r)$ is the RDF, which is a measure of the probability of finding a molecule at a distance $r$ from another molecule, and $y(r)\equiv g(r) e^{\phi(r)/k_BT}$ is the so-called cavity function. On the other hand, the static structure factor $S(q)$, which is related to the Fourier transform of $g(r)-1$ by
\begin{equation}
\label{b0}
    S(q)=1+\rho \int d \mathbf{r}\, e^{-\imath \mathbf{q}\cdot \mathbf{r}} [g(r)-1],
\end{equation}
where $\imath$ is the imaginary unit and $q$ is the wavenumber, also provides another connection between the structural and thermodynamic properties. In fact, the isothermal compressibility of the fluid, $\kappa_T=\rho^{-1}\left(\partial
\rho/\partial p\right)_{T}$ is directly related to the
long-wavelength limit of the structure factor, namely
\beq
\chi_T\equiv \rho k_{\textrm{B}}T\kappa_T=S(0),
\label{b2}
\eeq
where $\chi_T$ is the isothermal susceptibility. This leads to the compressibility route to the equation of state:
\beq
\label{c_route}
Z=\int_0^1 \frac{dt}{\chi_T(\rho t)}.
\eeq

It is also convenient at this stage to introduce the Laplace transform $G(s)$ of $rg(r)$, i.e.,
\begin{equation}
\label{b3}
G(s)=\int_0^\infty d r\, e^{-rs} rg(r).
\end{equation}
In terms of $G(s)$, the static structure factor is simply expressed as
\begin{equation}
\label{b1}
S(q)=1-2\pi\rho \left.\frac{G(s)-G(-s)}{s}\right|_{s=\imath q}.
\end{equation}

Let us now consider the piecewise constant potential of our fluid given by
\begin{equation}
\label{a1}
\phi(r)=\left\{
\begin{array}{ll}
\infty  ,& r<\sigma, \\
\epsilon_1  ,& \sigma<r<\lambda_1, \\
\epsilon_2  ,& \lambda_1<r<\lambda_2 , \\
\vdots&\vdots \\
\epsilon_n  ,& \lambda_{n-1}<r<\lambda_n , \\
0,&r>\lambda_n  .
\end{array}
\right.
\end{equation}
This potential is characterized by a hard core of diameter $\sigma$ and $n$ steps of  heights $\epsilon_j$ and widths $\la_j-\la_{j-1}$, where henceforth the conventions $\la_0=1$, $\epsilon_0=\infty$, and $\epsilon_{n+1}=0$ are understood.  Without loss of generality in what follows we will  take the hard-core diameter $\sigma=1$ as the length unit and  $|\epsilon_1|$ as the energy unit.

For the potential \eqref{a1}, the virial route \eqref{presMul} to the equation of state adopts the specially appealing form
\begin{equation}
\label{Zvir}
Z=1+4\eta \sum_{j=0}^{n}\lambda_j^3\Delta g(\lambda_j),
\end{equation}
where $\eta\equiv\frac{\pi}{6}\rho^*$ is the packing fraction ($\rho^*=\rho\sigma^3$ being the reduced density), $\Delta g(\lambda_j)\equiv g(\lambda_j^+)-g(\lambda_j^-)$ is the jump of the RDF at $r=\lambda_j$, and use has been made of the continuity of $y(r)$.

We focus now on  some exact mathematical properties of the function $G(s)$. More specifically, we introduce an auxiliary function $F(s)$ defined through
\beq
\label{b5}
G(s)=s\frac{F(s)e^{-s}}{1+12\eta F(s)e^{-s}},
\eeq
so that Laplace inversion of Eq.~(\ref{b5}) provides a useful representation of the RDF:
 \begin{equation}
 \label{b6}
 g(r)=r^{-1}\sum_{m=1}^\infty (-12\eta)^{m-1}f_m(r-m)\Theta(r-m),
 \end{equation}
where $f_m(r)=\mathcal{L}^{-1}\left[s[F(s)]^m\right]$ is the inverse Laplace transform of $s[F(s)]^m$ and $\Theta(r)$ is the Heaviside step function.
The exact behaviors for large $s$ and small $s$ of  the auxiliary function $F(s)$ are \cite{YS91,YS94,YHS96,SYH12,S16,SYH20}
\begin{subequations}
\begin{equation}
\label{b9}
F(s)\sim s^{-2},\quad s\rightarrow \infty,
\end{equation}
\bal
\label{b10}
F(s)=&-\frac{1}{12\eta}\left(1+s+\frac{1}{2}s^2+\frac{1+2\eta}{12\eta}s^3
+\frac{2+\eta}{24\eta}s^4\right)\nn
&+F_5 s^5+F_6 s^6+{\cal O}(s^7),
\eal
\end{subequations}
where the values of the coefficients $F_5$ and $F_6$ determine the value of $S(0)$, and hence of the isothermal compressibility, by
\beq
S(0)=3456\eta^3\left(F_5-F_6\right)-1+8\eta+2\eta^2.
\label{b4}
\eeq

Let us now decompose $F(s)$ as
\beq
F(s)=\sum_{j=0}^n R_j(s)e^{-(\lambda_j-1)s}.
\label{c0}
\eeq
This decomposition reflects the discontinuities of $g(r)$ at the points $r=\la_j$. More specifically,
\beq
\Delta g(\la_j)=\frac{1}{\la_j}\lim_{s\to\infty}s^2R_j(s)=\frac{1}{\la_j}\xi_j(0),\quad j=0,1,\ldots,n,
\label{disc2}
\eeq
where  $\xi_j(r)=\mathcal{L}^{-1}\left[s R_j(s)\right]$. Thus,
\beq
f_1(r)=\sum_{j=0}^n \xi_j(r-\lambda_j+1)\Theta(r-\lambda_j+1).
\label{c1}
\eeq

\subsection{Our approximation}

\subsubsection{RDF}

Now we assume the following \emph{rational-function} approximation for $R_j(s)$:
\beq
R_j(s)=-\frac{1}{12\eta}\frac{A_j+B_j s}{1+S_1 s+S_2 s^2+S_3 s^3}, \quad j=0,1,\ldots,n.
\label{c6}
\eeq
Since the degree difference between the numerator and denominator of $R_0(s)$ is equal to $2$, the form \eqref{c6} ensures the consistency with Eq.\ \eqref{b9}. Moreover, according to Eq.\ \eqref{disc2},
\beq
\Delta g(\lambda_j)=-\frac{1}{12\eta\la_j}\frac{B_j}{S_3},\quad j=0,1,\ldots,n.
\label{Xi1}
\eeq

For simplicity, the parameters $\{A_j,j=0,1,\ldots,n\}$ are assumed to be independent of density, yielding \cite{SYH12}
\beq
A_j=e^{-\epsilon_{j+1}/k_BT}-e^{-\epsilon_{j}/k_BT}.
\eeq
Next, the exact expansion (\ref{b10}) gives
\begin{subequations}
\label{c8-c11}
\beq
S_1=\Omega_0-\Lambda_1,
\label{c8}
\eeq
\beq
S_2=\frac{1}{2}\Lambda_2-\Omega_1,
\label{c9}
\eeq
\beq
S_3=\frac{1}{2}\Omega_2-\frac{1}{6}\Lambda_3-\frac{1}{12\e},
\label{c10}
\eeq
\beq
\Lambda_1+\frac{1}{2}\e\Lambda_4=\Omega_0+2\e\Omega_3,
\label{c11}
\eeq
\end{subequations}
where
\beq
\Lambda_\ell\equiv \sum_{j=0}^n A_j \lambda_j^\ell,\quad \Omega_\ell\equiv \sum_{j=0}^n B_j \lambda_j^\ell.
\label{c12}
\eeq
Equations \eqref{c8}--\eqref{c10} give $S_1$, $S_2$, and $S_3$ as linear combinations of the coefficients $\{B_j\}$, while the constraint \eqref{c11} allows us to express one of the latter coefficients, say $B_0$, in terms of the other ones.
To close the problem, we  then need $n$ additional constraints to determine $\{B_j\}$. The continuity of the cavity function imposes the conditions
\beq
\label{cond}
\xi_j(0)=e^{\epsilon_j/k_BT}A_j\sum_{i=0}^{j-1}\xi_i(\lambda_j-\lambda_i),\quad  j=1,\ldots,n.
\eeq
Application of the residue theorem gives
\beq
\xi_j(r)=-\frac{1}{12\eta}\sum_{\alpha=1}^3 \frac{A_j+B_j s_\alpha}{S_1 +2S_2 s_\alpha+3S_3 s_\alpha^2}s_\alpha e^{s_\alpha r},
\label{xi}
\eeq
where $s_\alpha$ ($\alpha=1,2,3$) are the three roots of the cubic equation $1+S_1 s_\alpha+S_2 s_\alpha^2+S_3 s_\alpha^3=0$.
Thus, taking into account Eqs.~\eqref{disc2} and \eqref{Xi1}, Eq.\ \eqref{cond} becomes
\bal
\frac{B_j}{S_3}=&e^{\epsilon_j/k_BT}A_j\sum_{\alpha=1}^3\frac{s_\alpha e^{\la_j s_\alpha}}{S_1 +2S_2 s_\alpha+3S_3 s_\alpha^2}\nn
&\times\sum_{i=0}^{j-1}(A_i+B_is_\alpha)e^{-\la_i s_\alpha},\quad  j=1,\ldots,n.
\label{c15_cc}
\eal
Equation \eqref{c15_cc}, complemented by Eqs.~\eqref{c8-c11}, must be solved numerically.  Once $F(s)$ is fully determined, Eq.~\eqref{b6} provides the RDF, where $f_m(r)$ can be obtained from the residue theorem or, alternatively, by numerical Laplace inversion  \cite{AW92}.

It can be proved \cite{YS94,SYH12,S16} that the RFA reduces to the exact solutions of the Percus--Yevick integral equation theory for hard spheres and sticky hard spheres in the appropriate limits.

\subsubsection{Equation of state}
In the RFA, the equation of state from the virial route is obtained by inserting Eq.~\eqref{Xi1} into Eq.~\eqref{Zvir}:
\begin{equation}
\label{ZvirRFA}
Z=1-\frac{ {\Omega}_2}{3S_3} .
\end{equation}

In the case of the compressibility route, we need to expand $F(s)$ in powers of $s$, identify the coefficients $F_5$ and $F_6$, and make
use of Eqs.~\eqref{b2} and \eqref{b4}. After some algebra, one finds
\bal
\label{ZcompRFA}
\chi_T=&1+4\eta\left(\Lambda_3-3\Lambda_1\Lambda_2+3\Lambda_2 {\Omega}_0
+6\Lambda_1 {\Omega}_1-6 {\Omega}_0 {\Omega}_1\right.\nn
&\left.-3 {\Omega}_2\right)+\frac{2}{5}\eta^2\left(\Lambda_6-6\Lambda_1\Lambda_5+6\Lambda_5 {\Omega}_0+30\Lambda_1 {\Omega}_4\right.\nn
&\left.
-30 {\Omega}_0 {\Omega}_4-6 {\Omega}_5\right).
\eal
The corresponding compressibility factor is obtained from Eq.~\eqref{c_route} by numerical integration.

\subsection{RDF to first order in density and third virial coefficient}
\label{appA}
For the sake of completeness, let us obtain the explicit expressions of the RFA in the low-density regime.
\subsubsection{RDF to first order in density}
In the low-density limit we can write
\beq
B_j=A_j\left(\lambda_j+\eta\XX_j\right)+\mathcal{O}(\eta^2),\quad  j=0,1,\ldots, n,
\eeq
where the first-order coefficients $\XX_j$ are to be determined. From Eqs.\ \eqref{c8-c11},
\begin{subequations}
\beq
\label{3}
A_0\XX_0=-\sum_{i=1}^nA_i \XX_i-\frac{3}{2}\Lambda_4,
\eeq
\beq
S_1=-\frac{3}{2}\eta \Lambda_4+\mathcal{O}(\eta^2),
\eeq
\beq
S_2=-\frac{1}{2} \Lambda_2-\e\sum_{j=0}^n A_j X_j\lambda_j+\mathcal{O}(\eta^2),
\eeq
\beq
S_3=-\frac{1}{12\eta}+\frac{1}{3} \Lambda_3+\frac{\eta}{2}\sum_{j=0}^n A_j X_j\lambda_j^2+\mathcal{O}(\eta^2).
\eeq
\end{subequations}
Therefore, Eq.~\eqref{c6} becomes
\beq
R_j(s)=R_j^{(0)}(s)+\eta R_j^{(1)}(s)+\mathcal{O}(\eta^2),
\eeq
where
\begin{subequations}
\beq
\frac{R_j^{(0)}(s)}{A_j}=s^{-3}+\lambda_j s^{-2},
\eeq
\beq
\frac{R_j^{(1)}(s)}{A_j}
= \XX_js^{-2}+ 12(s^{-1}+\lambda_j )\left(s^{-5}-\frac{s^{-3}\Lambda_2}{2}+\frac{s^{-2}\Lambda_3}{3}\right).
\eeq
\end{subequations}
This yields
\beq
\label{xi_j}
\xi_j(r)\equiv\mathcal{L}^{-1}\left[sR_j(s)\right]=\xi_j^{(0)}(r)+\eta \xi_j^{(1)}(r)+\mathcal{O}(\eta^2),
\eeq
with
\begin{subequations}
\label{xi_j01}
\beq
\label{xi_j0}
\frac{\xi_j^{(0)}(r)}{A_j}=\lambda_j+r,
\eeq
\beq
\label{xi_j1}
\frac{\xi_j^{(1)}(r)}{A_j}
= \XX_j+\frac{r^3}{2}(4\lambda_j+r)-3\Lambda_2r(2\lambda_j+r)+4\Lambda_3(\lambda_j+r).
\eeq
\end{subequations}

We still need to determine the coefficients $\{\XX_j, j=1,\ldots,n\}$. From the continuity  conditions \eqref{cond}, we get
\beq
\label{11}
\XX_j=e^{\epsilon_j/k_BT}\left(\sum_{i=0}^{j-1}A_i \XX_i+K_j\right),\quad j=1,\ldots,n,
\eeq
where
\beq
K_{j}\equiv \sum_{i=0}^{j-1}A_i\left[\frac{(\lambda_j-\lambda_i)^3}{2}(\lambda_j+3\lambda_i)-3\Lambda_2(\lambda_j^2-\lambda_i^2)\right].
\eeq
It can be checked that the solution to Eq.~\eqref{11} is
\beq
\label{solu}
\XX_j=\sum_{i=j+1}^n A_i^+ K_i+e^{\epsilon_{j+1}/k_BT}K_j-\frac{3}{2}\Lambda_4,\quad 1\leq j\leq n,
\eeq
where
\beq
A_j^+\equiv e^{\epsilon_{j+1}/k_BT}-e^{\epsilon_j/k_BT},\quad j=0,1,\ldots, n.
\eeq
Taking $K_0=0$, Eq.~\eqref{solu} can still be used to get $\XX_0$, thus obtaining an expression equivalent to Eq.~\eqref{3}.

Once we have obtained all the RFA parameters to first order in density, let us evaluate the RDF to that order.
Recalling Eqs.~\eqref{b5} and \eqref{c0}, we get
\beq
G(s)=G^{(0)}(s)+\eta G^{(1)}(s)+\mathcal{O}(\eta^2),
\eeq
where
\begin{subequations}
\beq
G^{(0)}(s)=sF^{(0)}(s)e^{-s},
\eeq
\beq
G^{(1)}(s)=sF^{(1)}(s)e^{-s}-12 s \left[F^{(0)}(s)\right]^2e^{-2s},
\eeq
\end{subequations}
with
\begin{subequations}
\beq
F^{(0)}(s)\equiv \sum_{j=0}^n R_j^{(0)}(s)e^{-(\lambda_j-1)s},
\eeq
\beq
F^{(1)}(s)\equiv \sum_{j=0}^n R_j^{(1)}(s)e^{-(\lambda_j-1)s}.
\eeq
\end{subequations}
Thus,
\beq
\label{g0g1}
g(r)=g^{(0)}(r)+\eta g^{(1)}(r)+\mathcal{O}(\eta^2),
\eeq
where
\begin{subequations}
\label{g0-g1}
\bal
g^{(0)}(r)=&r^{-1}\sum_{j=0}^n \xi^{(0)}(r-\lambda_j)\Theta(r-\lambda_j)\nn
=&\sum_{j=0}^n A_j\Theta(r-\lambda_j),
\eal
\bal
g^{(1)}(r)=&r^{-1}\sum_{j=0}^n \xi^{(1)}(r-\lambda_j)\Theta(r-\lambda_j)\nn
&-12\eta r^{-1}f_2^{(0)}(r-2)\Theta(r-2),
\eal
\end{subequations}
$f_2^{(0)}(r)$ being the inverse Laplace transform of $s \left[F^{(0)}(s)\right]^2$.
Using
\bal
s \left[F^{(0)}(s)\right]^2=& \sum_{i=0}^n\sum_{j=0}^n A_i A_j\left[s^{-5}+(\lambda_i+\lambda_j) s^{-4}\right.\nn
&\left.+\lambda_i\lambda_js^{-3}\right]e^{-(\lambda_i+\lambda_j-2)s},
\eal
we get
\bal
\label{f20}
f_2^{(0)}(r-2)=& \sum_{i=0}^n\sum_{j=0}^n A_i A_j\frac{(r-\lambda_i-\lambda_j)^2}{24}\left[(r-\lambda_i-\lambda_j)^2\right.\nn
&\left.+4(\lambda_i+\lambda_j) (r-\lambda_i-\lambda_j)+12\lambda_i\lambda_j\right]\nn
&\times\Theta(r-\lambda_i-\lambda_j).
\eal
It can be checked that
\begin{subequations}
\beq
g^{(0)}(r)=1,\quad r\geq \lambda_n,
\eeq
\beq
g^{(1)}(r)=0,\quad r\geq 2\lambda_n.
\eeq
\end{subequations}

Equations \eqref{g0g1}, \eqref{g0-g1}, and \eqref{f20} provide the RDF to first order in density within the RFA. While $g^{(0)}(r)$ is exact, $g^{(1)}(r)$ is approximate. From standard statistical-mechanical tools \cite{S16,HM13}, the exact result can be written as
\begin{subequations}
\beq
g^{(1)}_{\text{exact}}(r)=e^{-\phi(r)/k_BT}y^{(1)}_{\text{exact}}(r),
\eeq
\beq
y^{(1)}_{\text{exact}}(r)=\frac{3}{\pi^3}\int_0^\infty dq\, q^2 \frac{\sin(qr)}{qr}\left[\widetilde{f}(q)\right]^2,
\eeq
\end{subequations}
where
\bal
\widetilde{f}(q)=&4\pi \int_0^\infty dr\, r^2 \frac{\sin(qr)}{qr}\left[e^{-\phi(r)/k_BT}-1\right]\nn
=&4\pi\sum_{j=0}^n A_j\frac{q\lambda_j\cos(q\lambda_j)-\sin(q\lambda_j)}{q^3}
\eal
is the Fourier transform of the Mayer function.
The $q$-integration can be performed to get
\bal
\label{y1exact}
y^{(1)}_{\text{exact}}(r\geq 1)=&\frac{1}{2}\sum_{i=0}^n\sum_{j=0}^n A_i A_j \frac{(\lambda_i+\lambda_j-r)^2}{r}\nn
&\times[(r+\lambda_i+\lambda_j)^2 - 4(\lambda_i^2+\lambda_j^2-\lambda_i\lambda_j)]\nn
&\times\Theta(\lambda_i+\lambda_j-r).
\eal
Note that, since $\lambda_j>1$, one always has $\lambda_i+\lambda_j>r$ if $r\leq 2$, so that the term $\Theta(\lambda_i+\lambda_j-r)$ is not needed in that case.
It can be checked that the RDF function $g^{(1)}(r)$ differs from $g^{(1)}_{\text{exact}}(r)$ in the interval $1\leq r\leq \lambda_n\leq 2$ only.

\begin{table*}[t]
\caption{\label{table:acronym} Systems examined in this paper.}
\begin{center}
\begin{tabular}{ccccccccccc}
\hline\hline
& & \multicolumn{2}{c}{SW parameters}&&\multicolumn{2}{c}{SS parameters}&&\multicolumn{2}{c}{SW parameters}\\
\cline{3-4} \cline{6-7} \cline{9-10}
 Label &$n$& $\lambda_1$& $\epsilon_1$ && $\lambda_2$& $\epsilon_2$ && $\lambda_3$& $\epsilon_3$  &Graph \\
\hline
A&$1$&$1.15$&$-1$&&&&&&&\includegraphics[width=\kk]{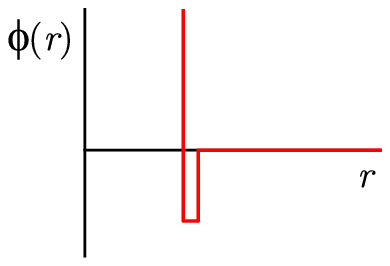}\\
\hdashline
B1&$2$&$1.15$&$-1$&&$1.25$&$0.25$&&&&\includegraphics[width=\kk]{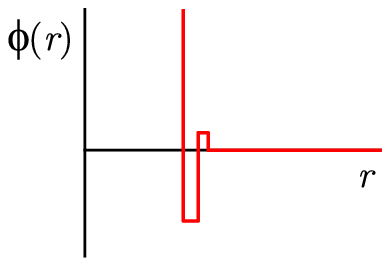}\\
\hdashline
B2&$2$&$1.15$&$-1$&&$1.25$&$1.00$&&&&\includegraphics[width=\kk]{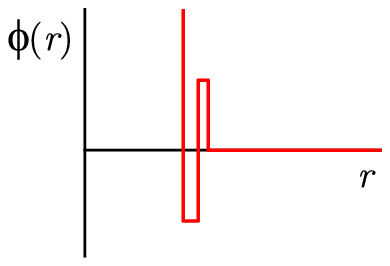}\\
\hdashline
B3&$2$&$1.15$&$-1$&&$1.50$&$0.25$&&&&\includegraphics[width=\kk]{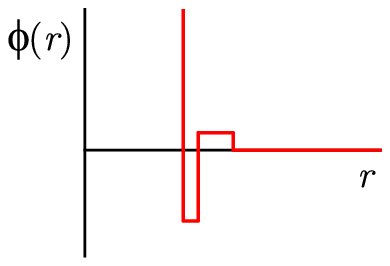}\\
\hdashline
B4&$2$&$1.15$&$-1$&&$1.50$&$1.00$&&&&\includegraphics[width=\kk]{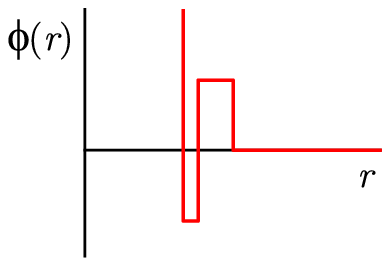}\\
\hdashline
C1&$3$&$1.15$&$-1$&&$1.50$&$0.50$&&$2$&$-0.1$&\includegraphics[width=\kk]{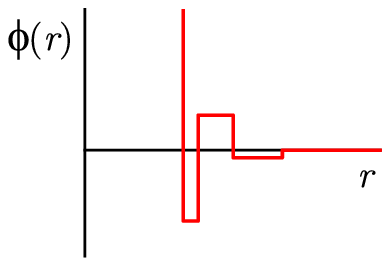}\\
\hdashline
C2&$3$&$1.15$&$-1$&&$1.50$&$0.50$&&$2$&$-0.2$&\includegraphics[width=\kk]{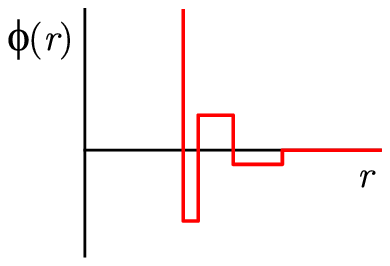}\\
\hline\hline
\end{tabular}
\end{center}
\end{table*}

\subsubsection{Second and third virial coefficients}

According to the virial route, Eq.~\eqref{Zvir},
\bal
Z=&1+4\eta\sum_{j=0}^n \lambda_j^2\xi_j(0)\nn
=&1+b_2\eta+b_{3,v}\eta^2+\mathcal{O}(\eta^3).
\eal
From Eqs.~\eqref{xi_j}  and \eqref{xi_j01} or, equivalently, from Eq.~\eqref{ZvirRFA}, we can identify
\beq
\label{bb2}
b_2=4\Lambda_3,
\eeq
\beq
b_{3,v}=16\Lambda_3^2+4\sum_{j=0}^n A_j X_j \lambda_j^2.
\eeq

Now we consider the compressibility route, Eqs.~\eqref{b2} and \eqref{c_route},
\bal
\chi_T=&1+24\eta\int_0^\infty dr\,r^2 \left[g(r)-1\right]\nn
=&1-2b_2\eta+(4b_2^2-3b_{3,c})\eta^2+\mathcal{O}(\eta^3),
\eal
where
\beq
b_{3,c}=\frac{64}{3}\Lambda_3^2-8\int_0^{2\lambda_n}dr\, r^2 g^{(1)}(r).
\eeq
An equivalent, but more explicit, closed expression for $b_{3,c}$ is obtained from Eq.~\eqref{ZcompRFA}:
\beq
b_{3,c}=\frac{64}{3}\Lambda_3^2-6\Lambda_2\Lambda_4+\frac{2}{3}\Lambda_6+4\sum_{j=0}^n  A_j\XX_j \lambda_j^2.
\eeq
The difference between $b_{3,c}$ and $b_{3,v}$ is
\beq
b_{3,c}-b_{3,v}=\frac{2}{3}\left(8\Lambda_3^2+\Lambda_6-9\Lambda_2\Lambda_4\right).
\eeq
This lack of thermodynamic consistency between the virial and compressibility routes is an expected consequence of the approximate character of the RFA.

Taking into account the property $\Delta g(\lambda_j)=A_j y(\lambda_j)$, the exact result is
\beq
b_{3,\text{exact}}=4\sum_{j=0}^n \lambda_j^3 A_j y^{(1)}_{\text{exact}}(\lambda_j),
\eeq
where $y^{(1)}_{\text{exact}}(r)$ is given by Eq.~\eqref{y1exact}.

\section{Illustrative results}
\label{results}

In this section we illustrate the results of our approach by examining their performance against recent simulation data \cite{PMPSGLVTC22} for both the structural and thermodynamic properties of this kind of fluids. For this purpose, it is convenient to order the different systems according to the number of steps after the hard core, so that the label A will indicate a single SW, the label B a SW followed by a SS, and the label C a SW followed by a SS and a second SW. According to the above nomenclature,  Table \ref{table:acronym} includes the values of the parameters $\{\lambda_j,\epsilon_j\}$ for each one of the seven examined systems.

\begin{figure*}%[tbp]
\begin{center}
\includegraphics[height=\hh]{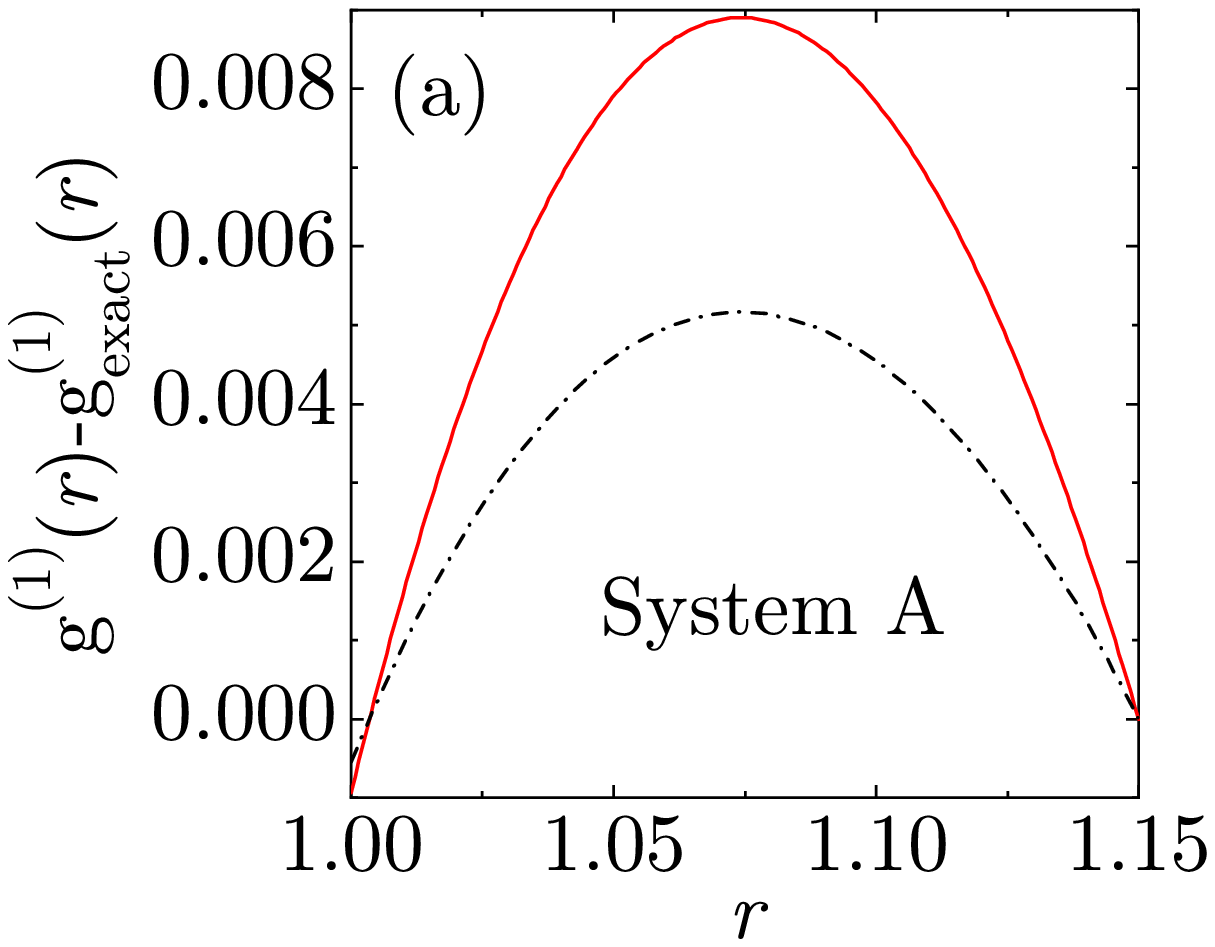}\hspace{1mm}\includegraphics[height=\hh]{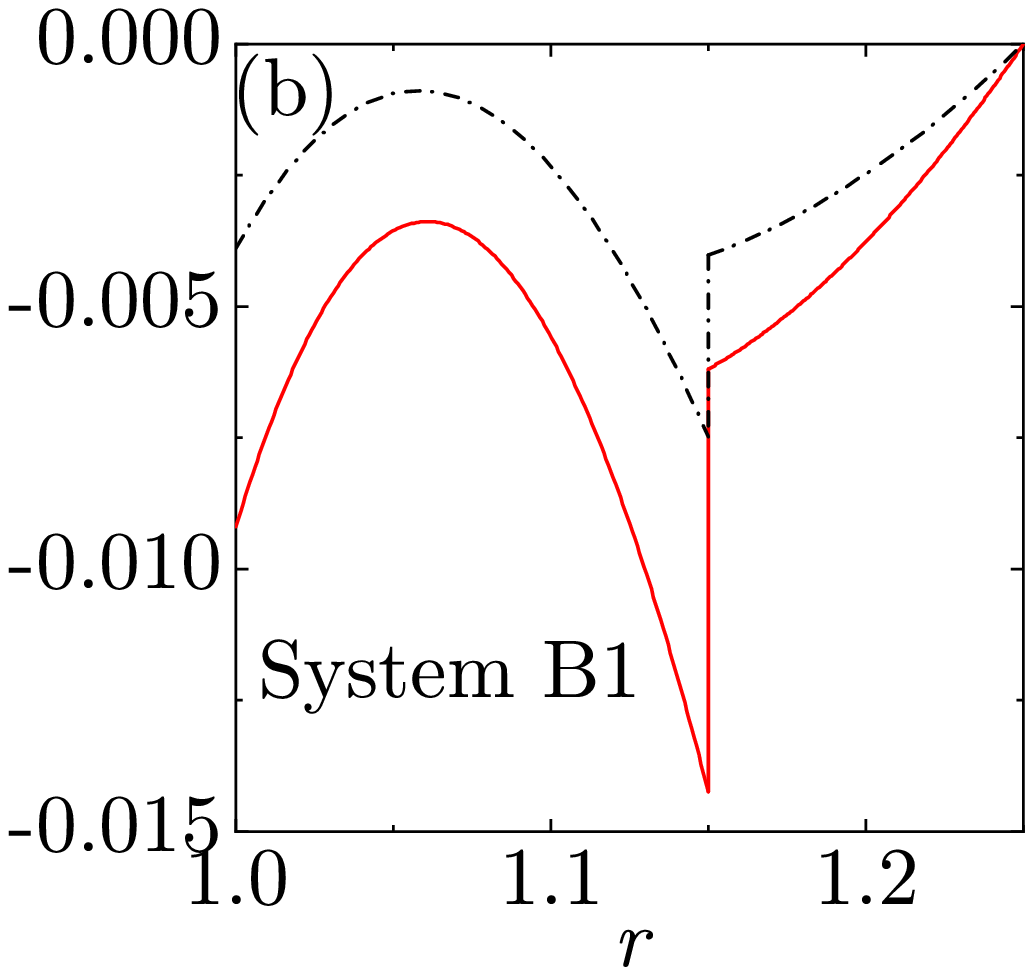}\hspace{1mm}\includegraphics[height=\hh]{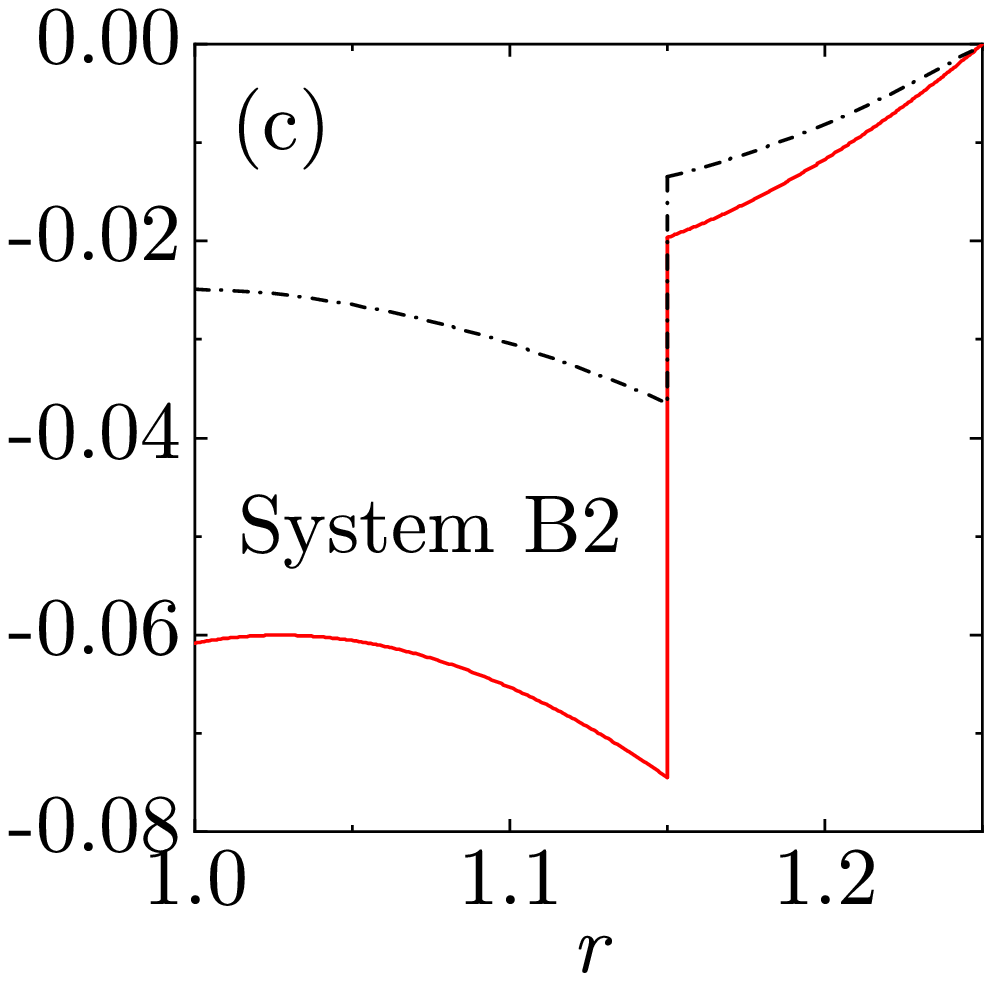}\hspace{1mm}\includegraphics[height=\hh]{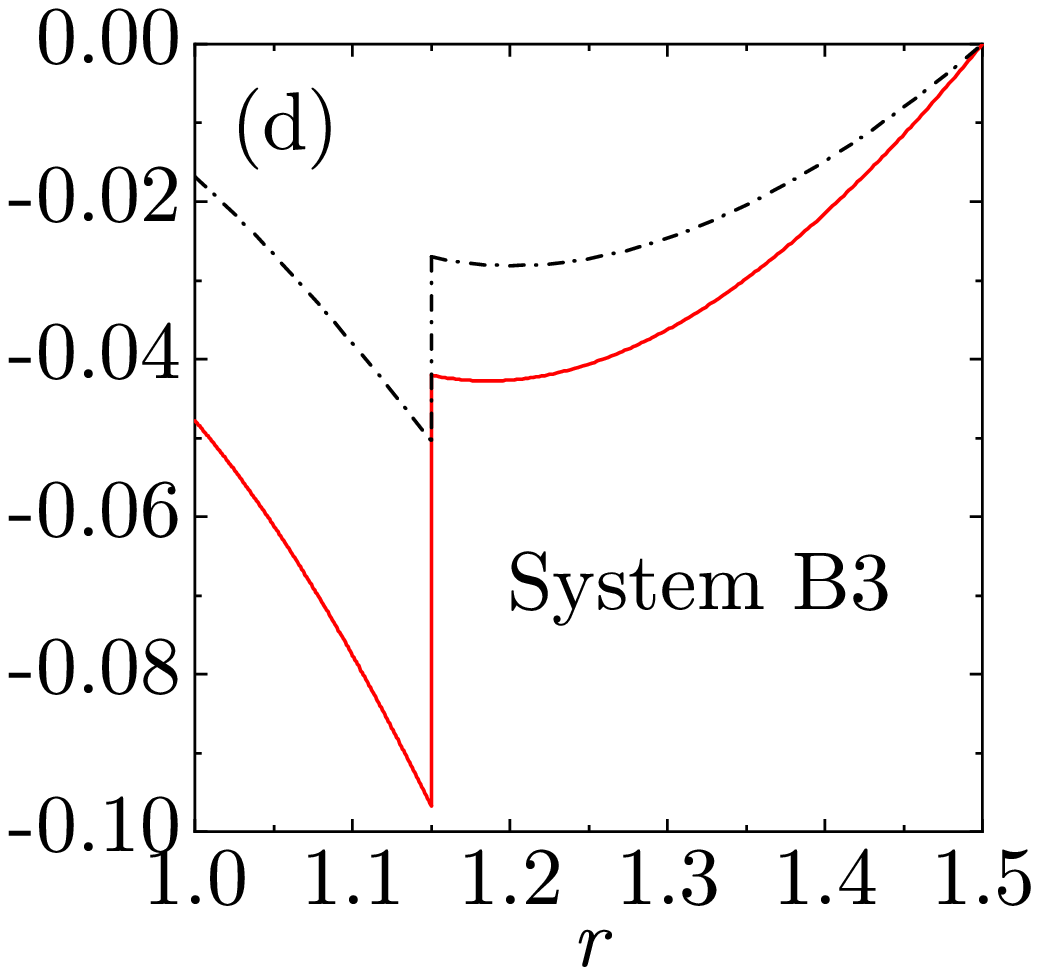}\\
\vspace{2mm}
\includegraphics[height=\hh]{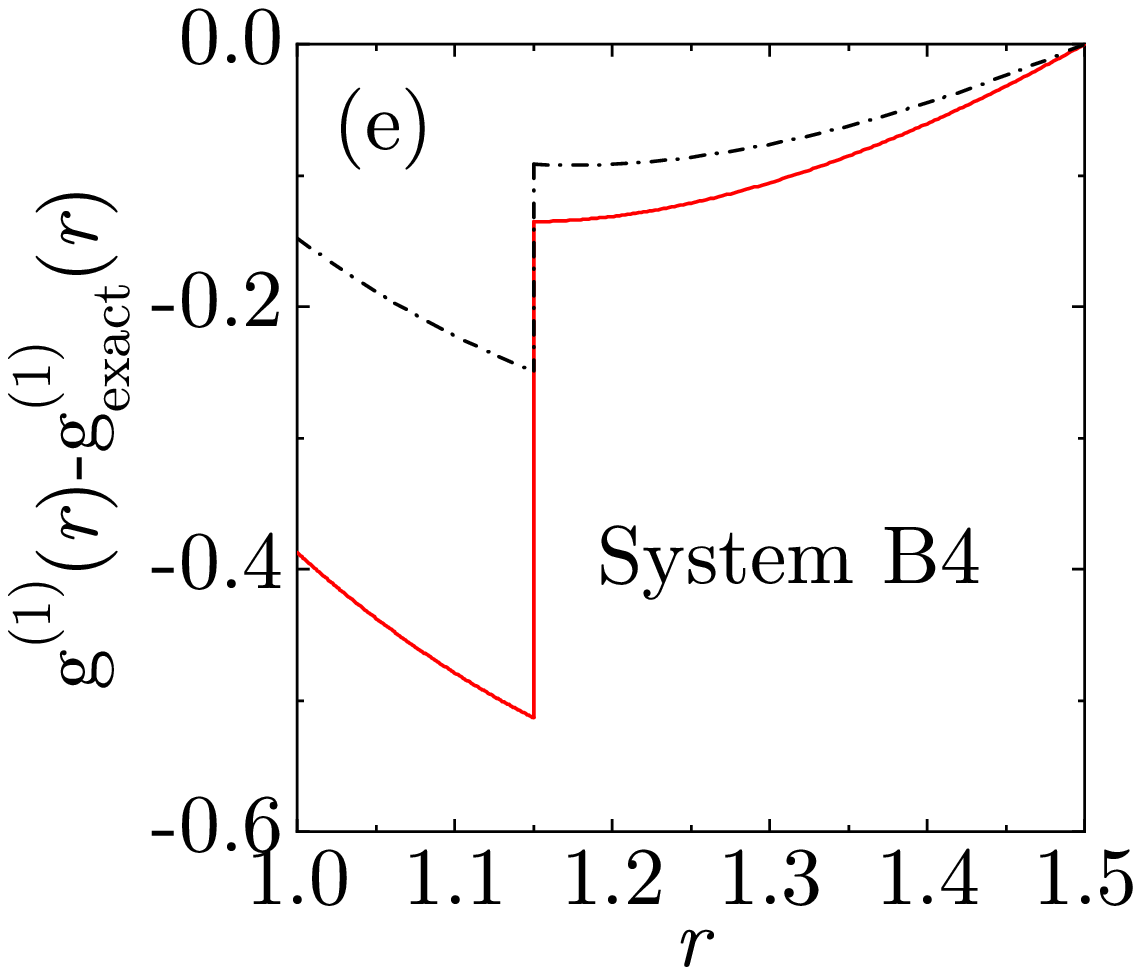}\hspace{1mm}\includegraphics[height=\hh]{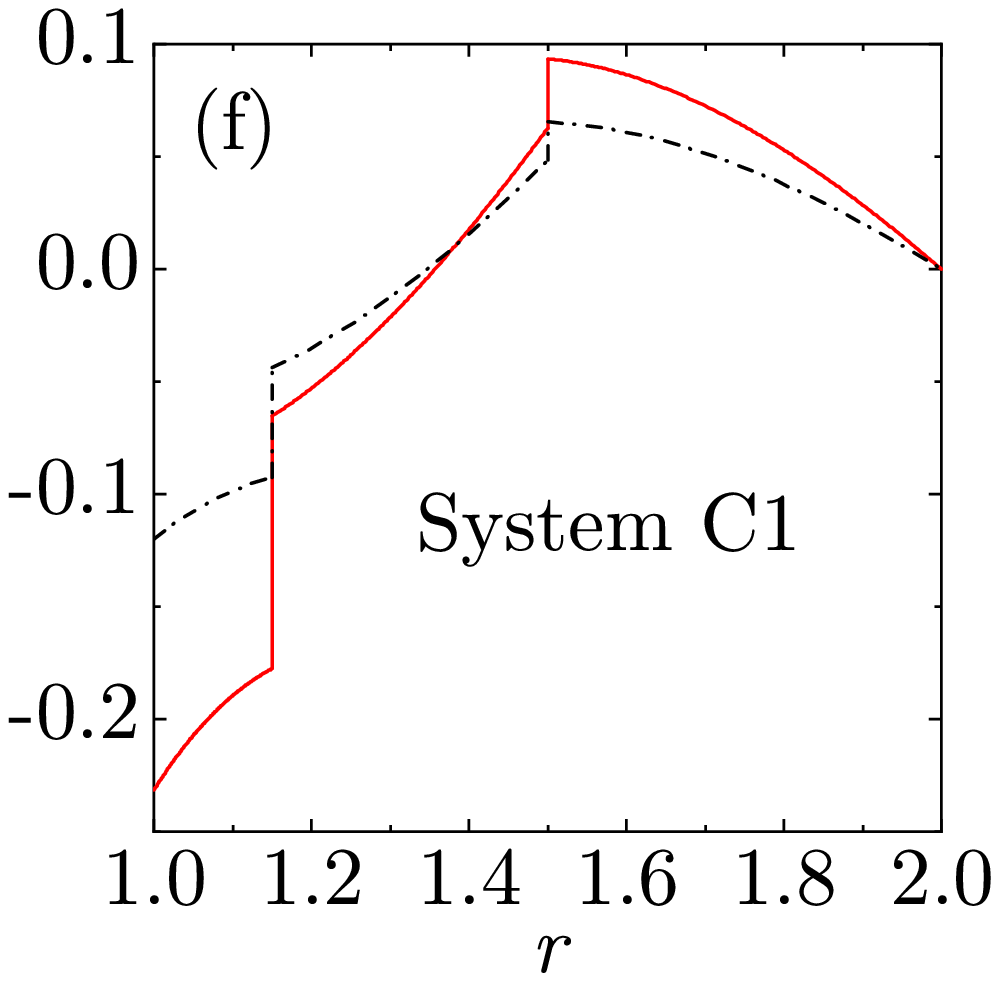}\hspace{1mm}\includegraphics[height=\hh]{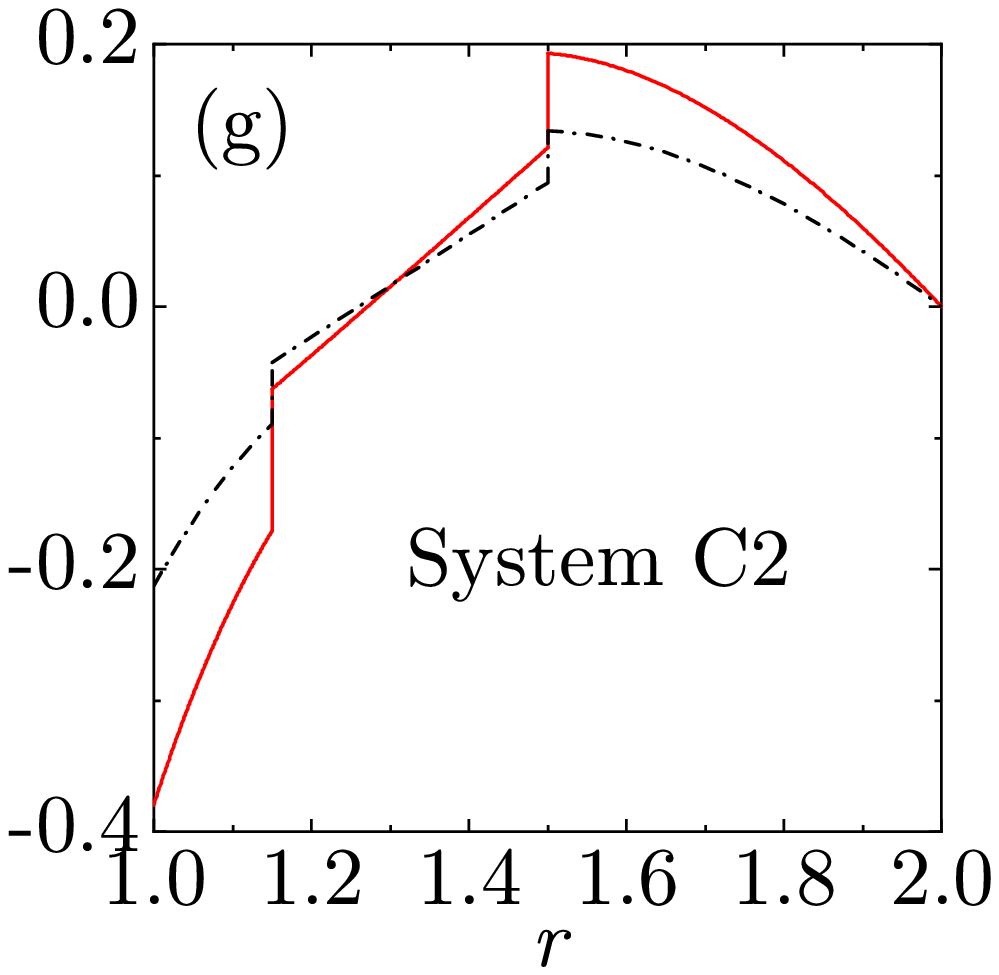}
\end{center}
 \caption{\label{fig:g1} Difference $g^{(1)}(r)-g^{(1)}_{\text{exact}}(r)$ at $T^=1.5$ (solid lines) and $T^*=2$ (dash-dotted lines) for the systems described in Table \ref{table:acronym}.}
 \end{figure*}

\begin{figure*}%[tbp]
\begin{center}
\includegraphics[height=\hhb]{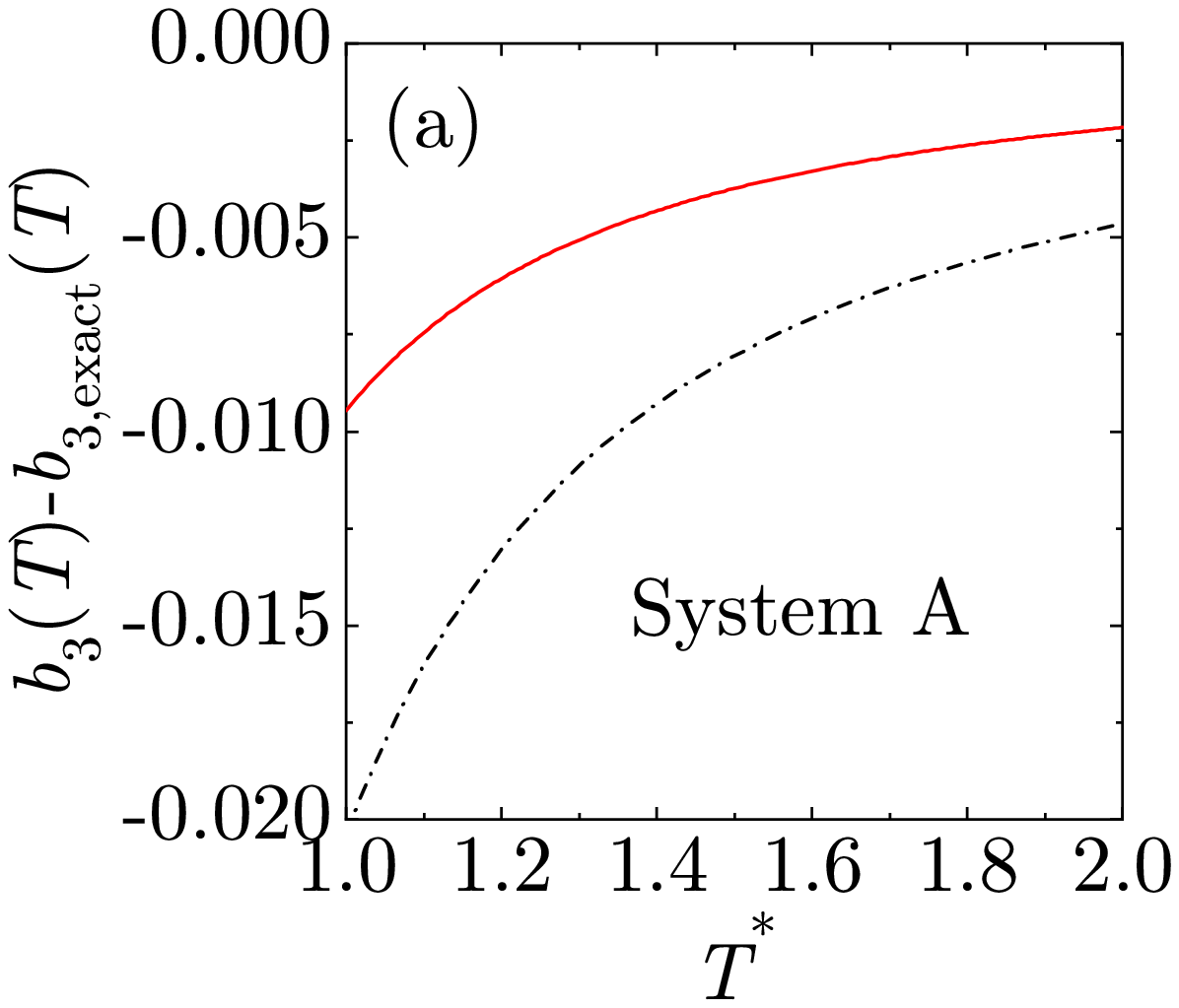}\hspace{1mm}\includegraphics[height=\hhb]{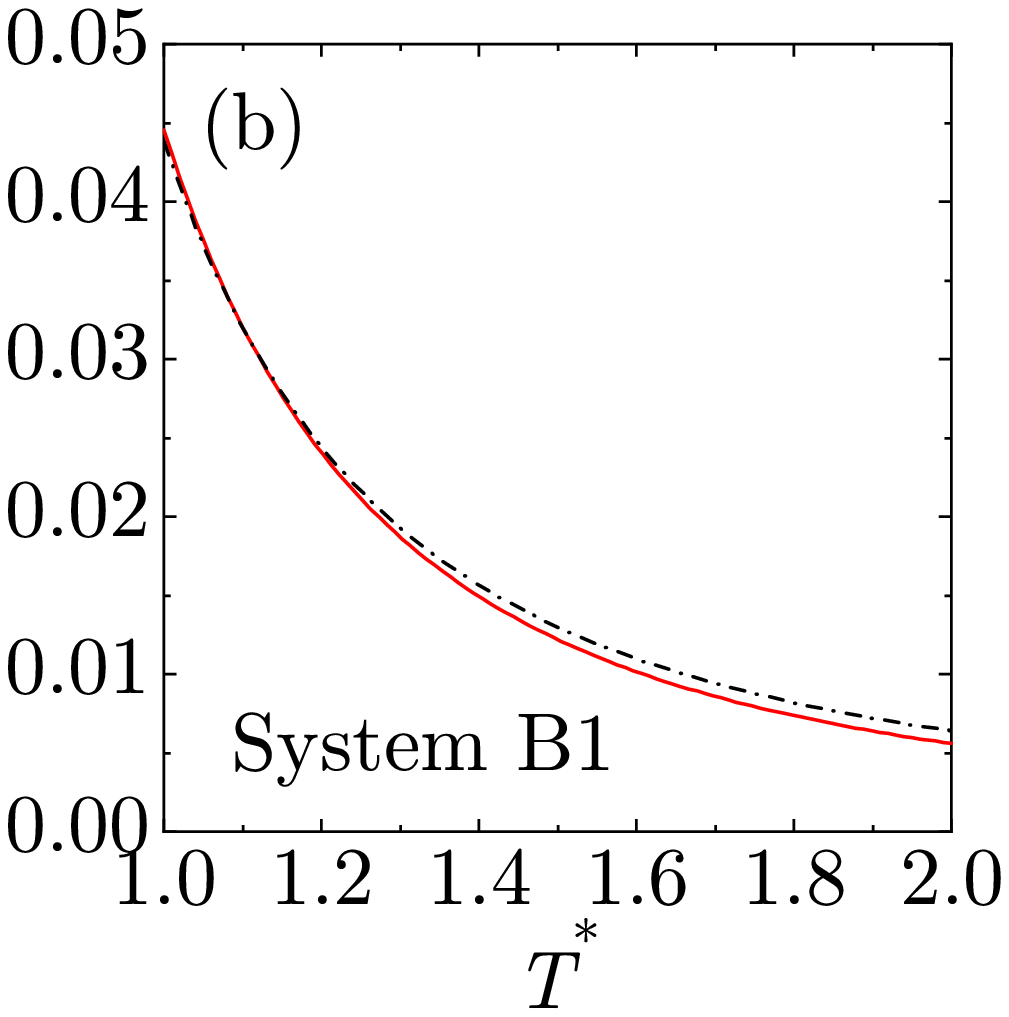}\hspace{1mm}\includegraphics[height=\hhb]{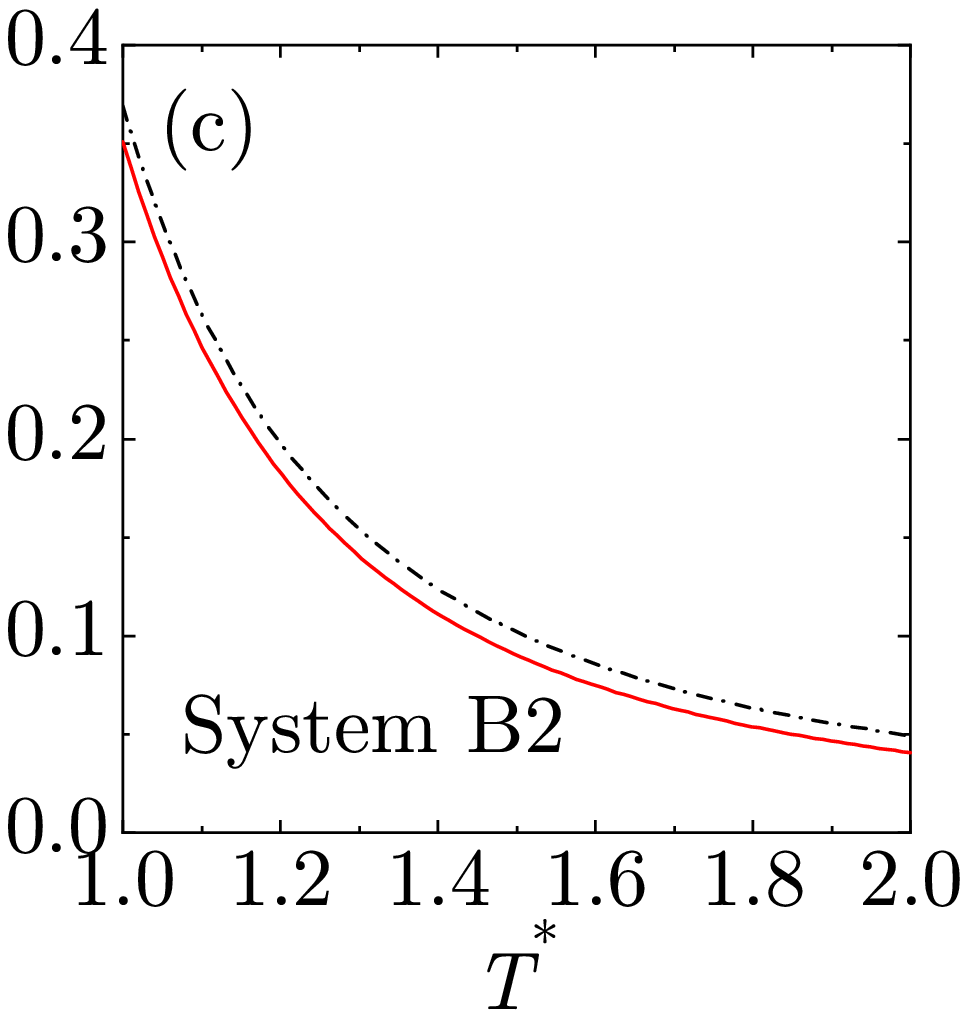}\hspace{1mm}\includegraphics[height=\hhb]{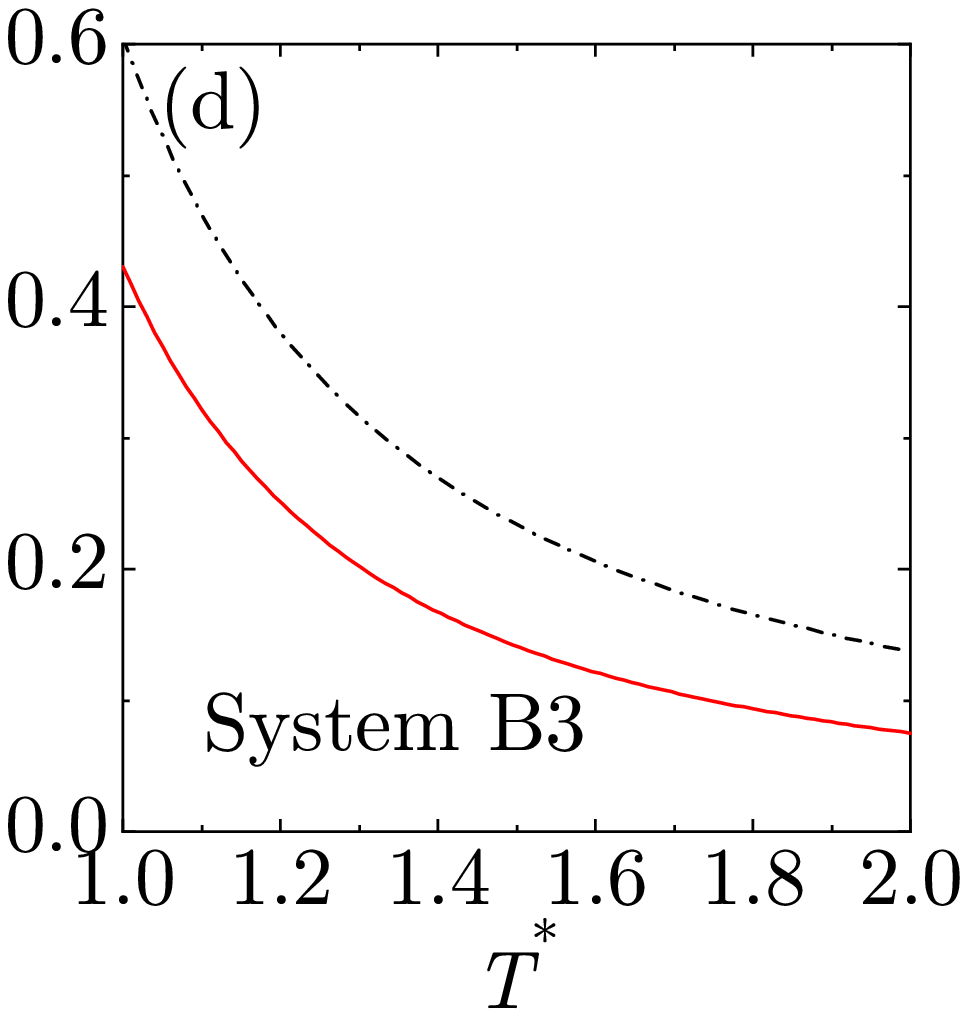}\\
\vspace{2mm}
\includegraphics[height=\hhb]{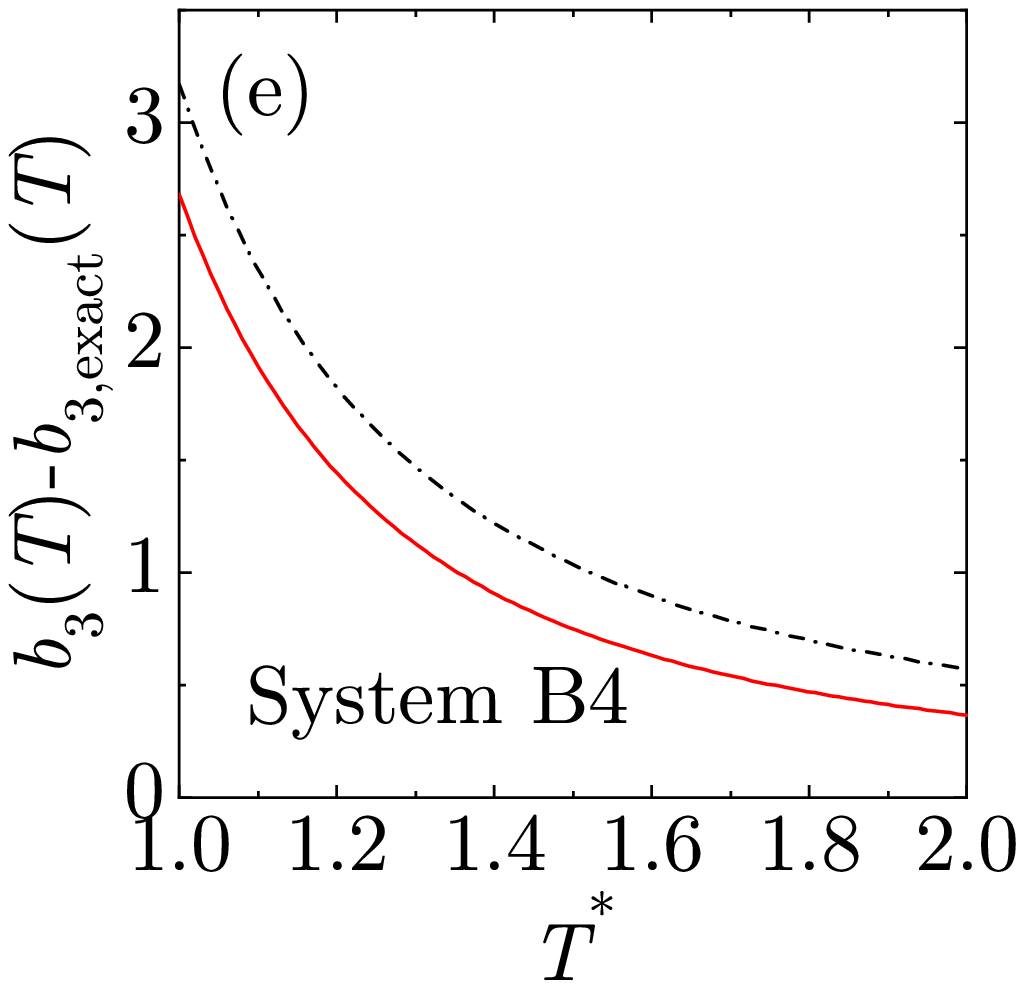}\hspace{1mm}\includegraphics[height=\hhb]{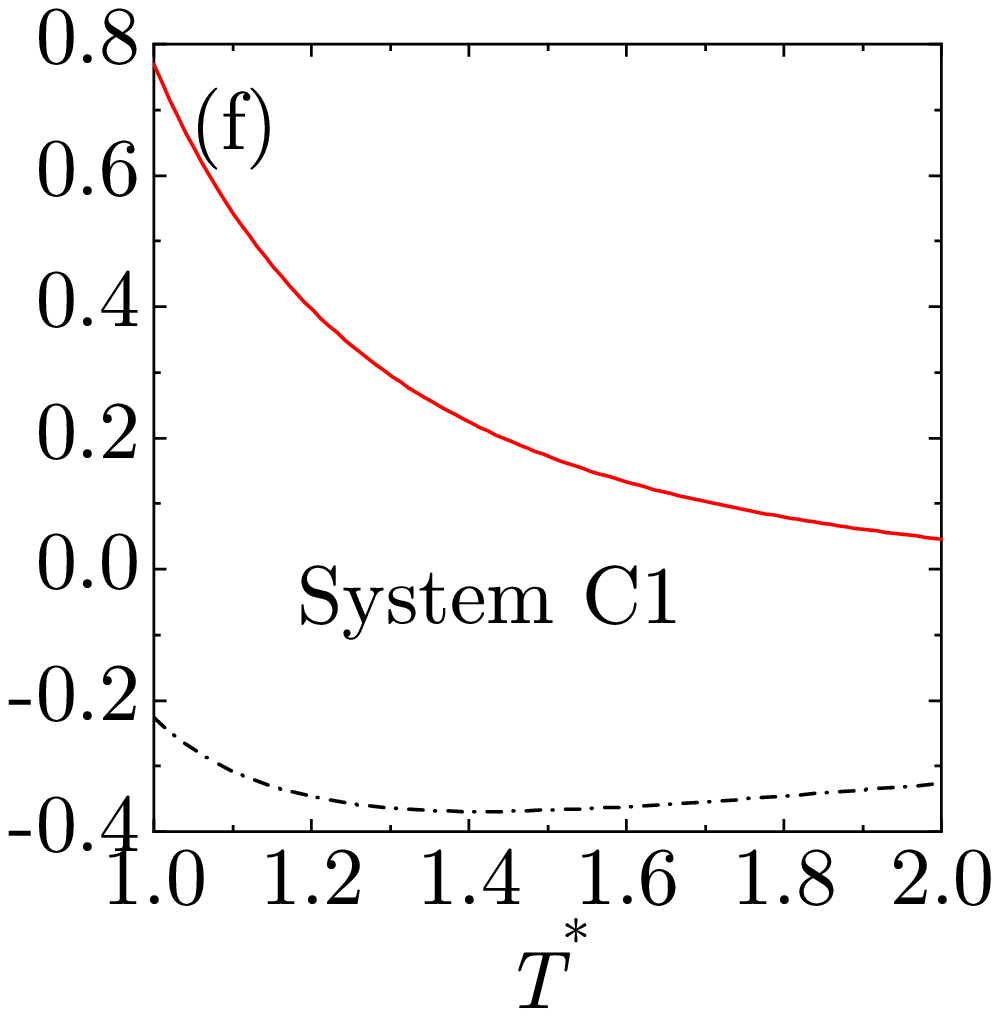}\hspace{1mm}\includegraphics[height=\hhb]{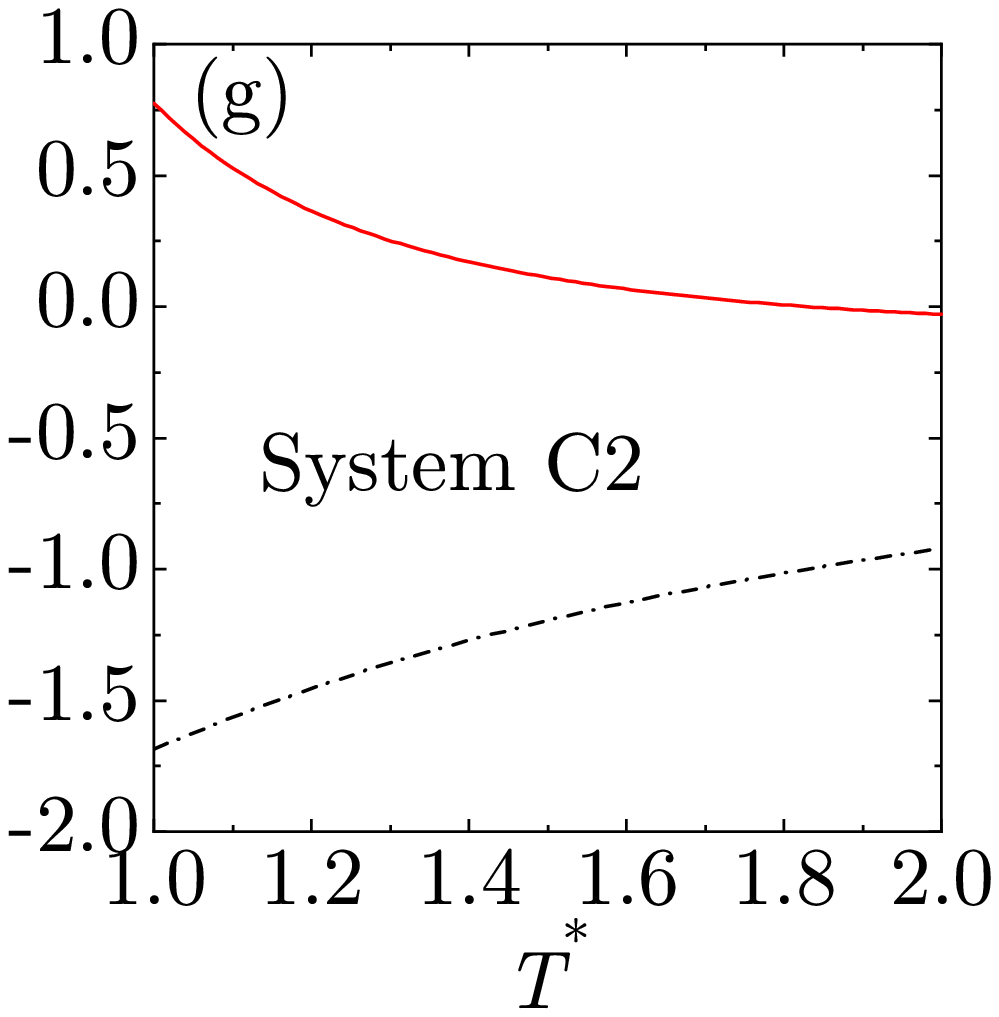}
\end{center}
 \caption{\label{fig:b3} Differences $b_{3,v}(T^*)-b_{3,\text{exact}}(T^*)$ (solid lines) and $b_{3,c}(T^*)-b_{3,\text{exact}}(T^*)$  (dash-dotted lines) for the systems described in Table \ref{table:acronym}.}
 \end{figure*}

\begin{figure*}%[tbp]
\begin{center}
\includegraphics[width=0.5\textwidth]{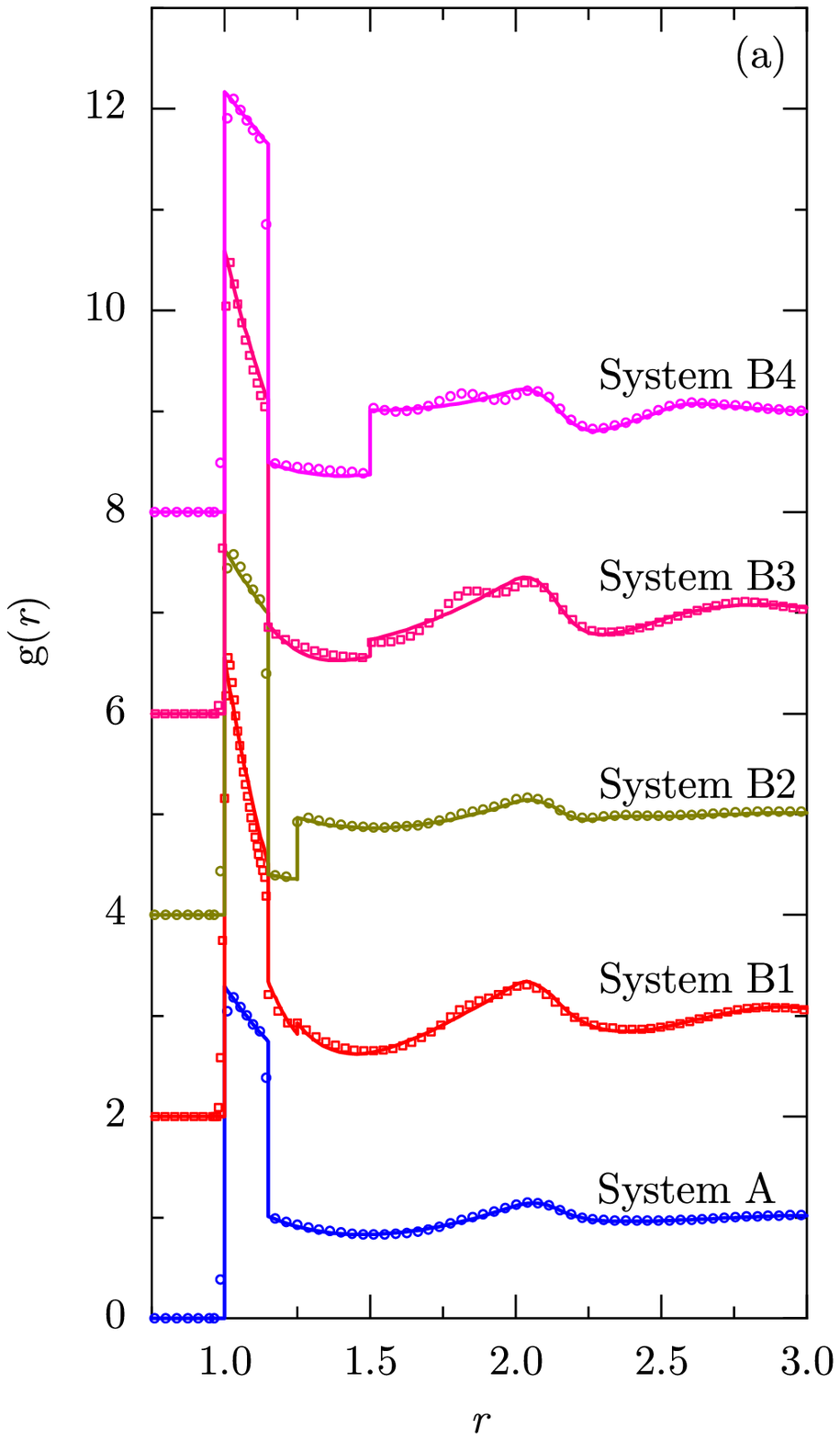}\includegraphics[width=0.5\textwidth]{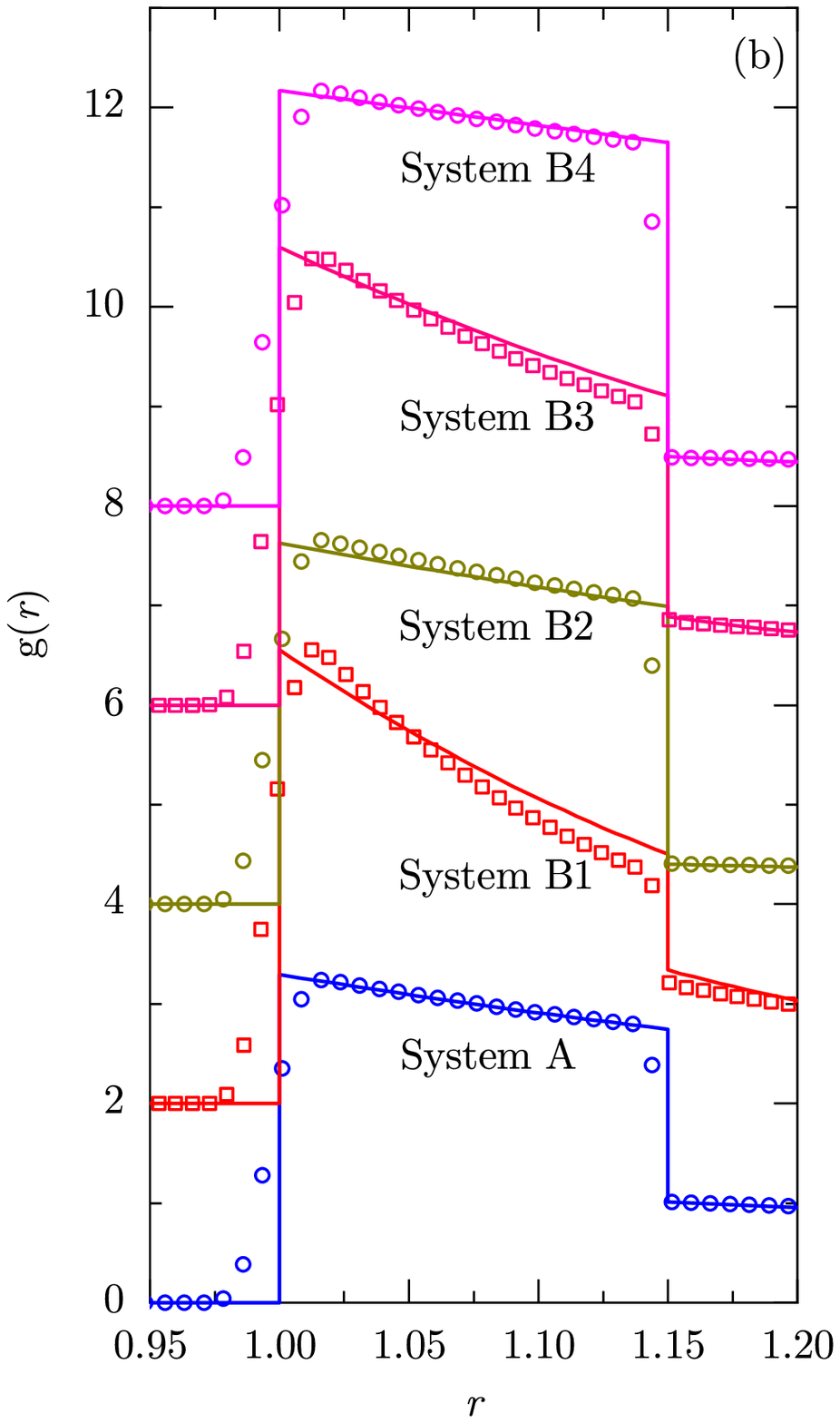}
\end{center}
 \caption{\label{fig:g_A-B} Comparison between the RDF as derived from the RFA  (solid lines) and simulation data \cite{PMPSGLVTC22} (symbols) for systems A--B4 with $(T^*,\rho^*)=(1,0.6)$ (systems A, B2, and B4), $(2,0.9)$ (system B1), and $(1,0.9)$ (system B3). Panel (a) shows the region $0.75\leq r\leq 3$, while panel (b) magnifies the region $0.95\leq r\leq 1.20$. For better clarity, the data have been shifted  upward  two, four, six, and eight units for systems B1--B4, respectively.}
 \end{figure*}

\begin{figure*}%[tbp]
\begin{center}
\includegraphics[width=0.5\textwidth]{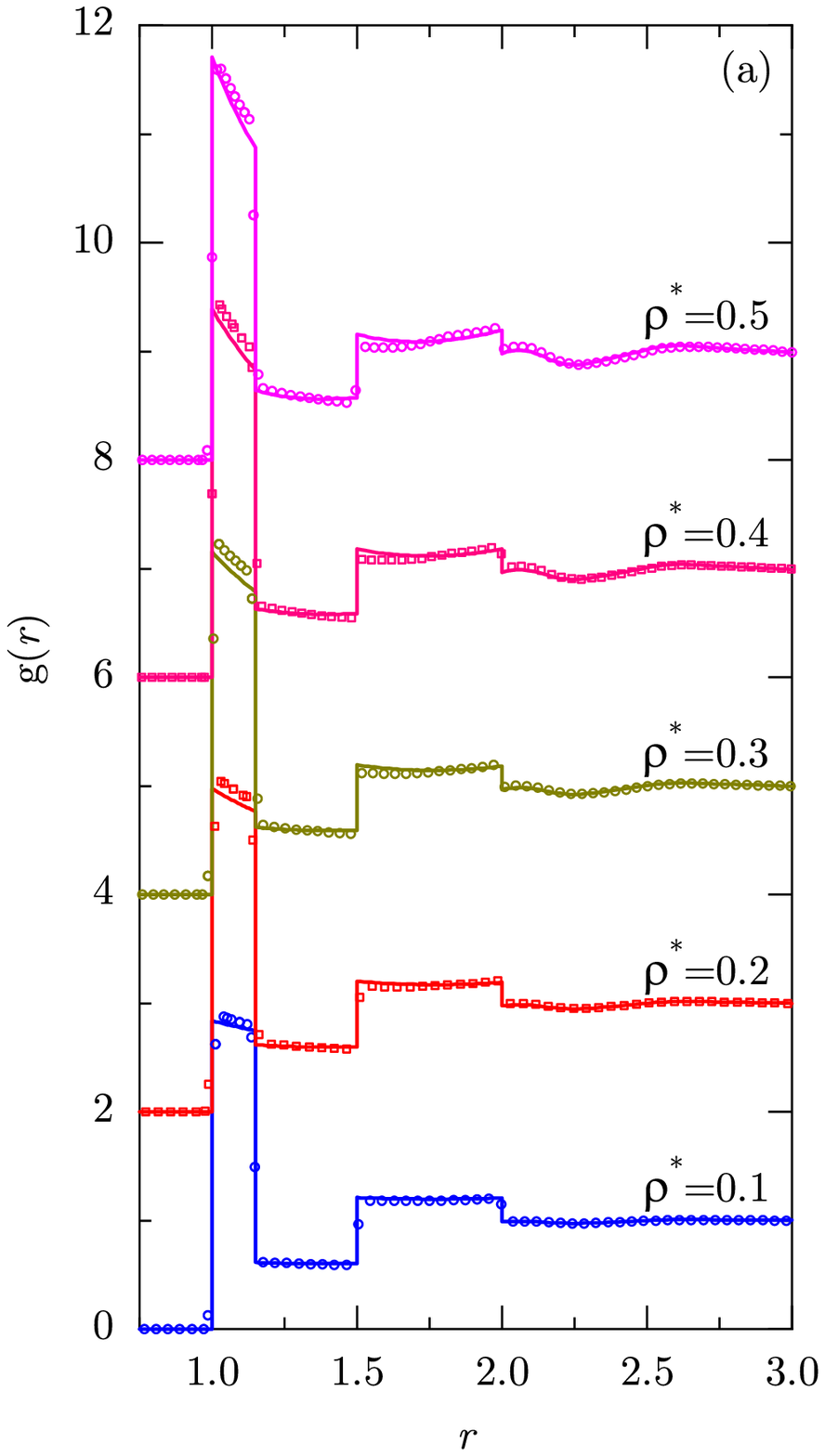}\includegraphics[width=0.5\textwidth]{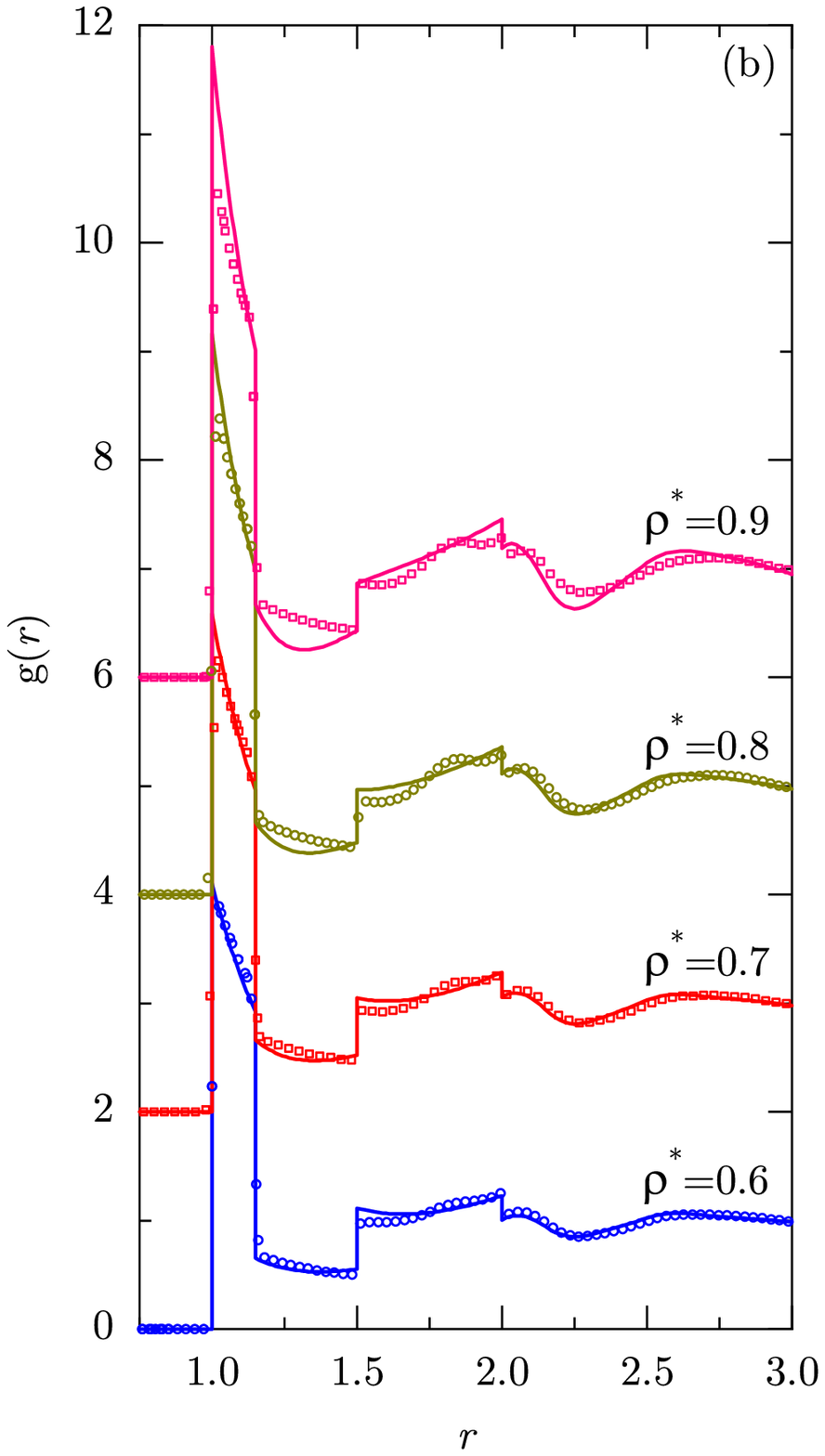}
\end{center}
 \caption{\label{fig:g_C} Comparison between the RDF as derived from the RFA  (solid lines) and simulation data \cite{PMPSGLVTC22} (symbols) for system C2 with $T^*=1$ and (a) $\rho^*=0.1,\ldots, 0.5$, (b) $\rho^*=0.6,\ldots, 0.9$. For better clarity, in each panel the data have been successively shifted  upward  two units.}
 \end{figure*}

  \begin{figure}%[tbp]
\begin{center}
\includegraphics[width=0.5\textwidth]{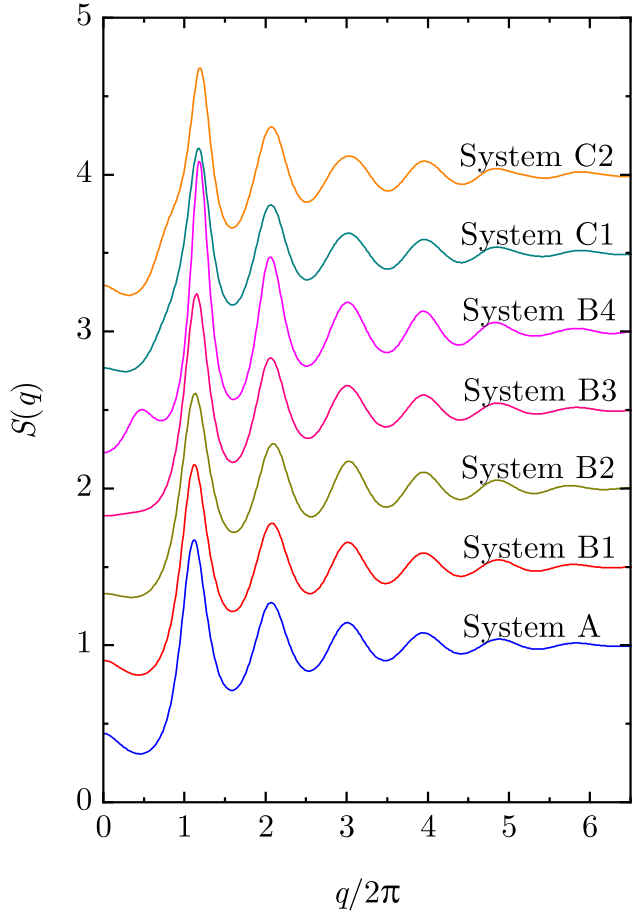}
\end{center}
 \caption{\label{fig:S_k}
 Plot of the structure factor $S(q)$ as derived from the RFA for systems A--C2 with  $(T^*,\rho)=(0.7,0.6)$. For better clarity,  the data have been successively shifted  upward  half a unit.}
 \end{figure}

 \begin{figure}%[tbp]
\begin{center}
\includegraphics[width=0.5\textwidth]{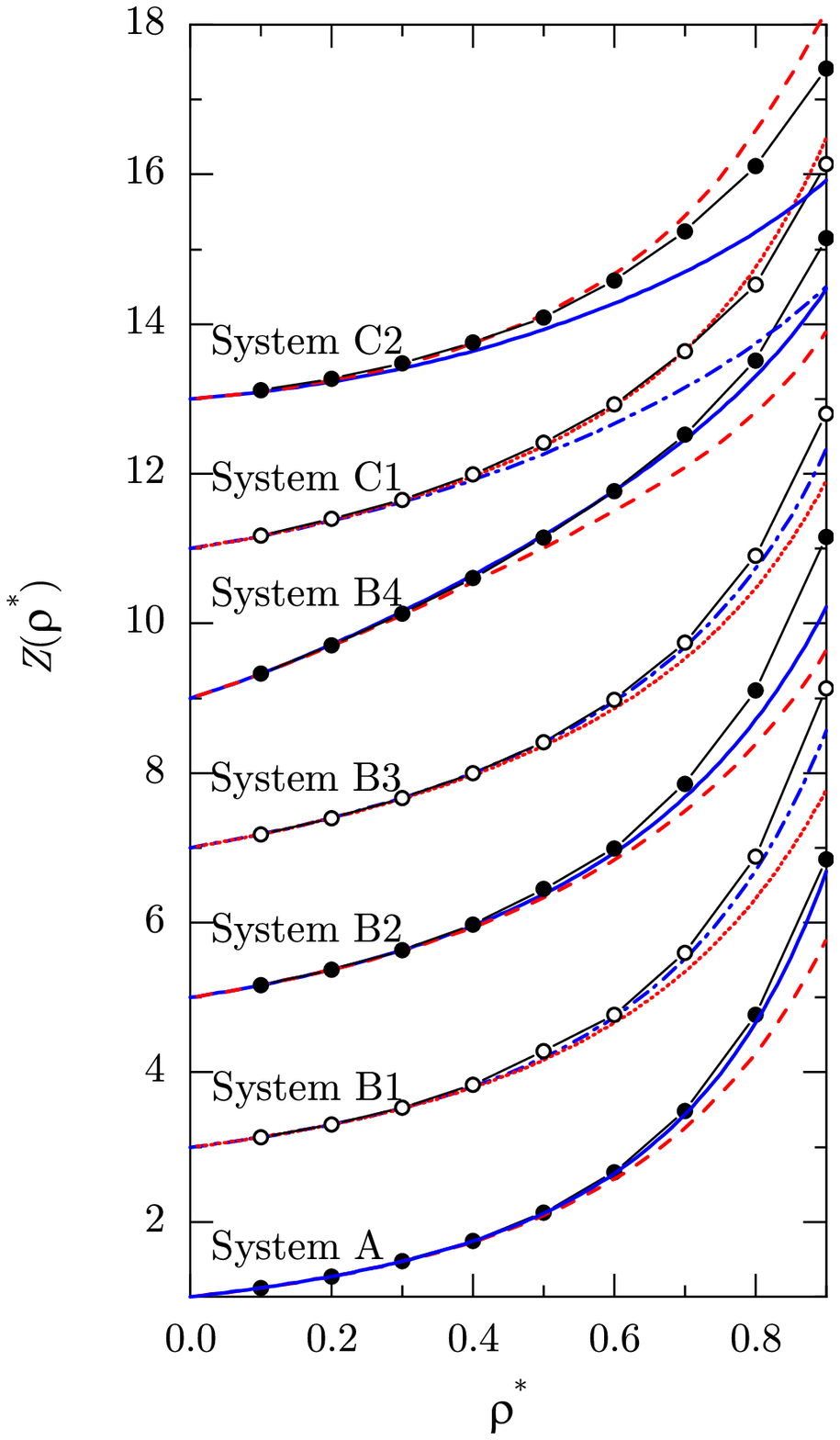}
\end{center}
 \caption{\label{fig:Z}
 Comparison between the density-dependence of the compressibility factor as derived from the RFA (virial route: dashed and dotted lines, compressibility route: solid and dash-dotted lines) and simulation data \cite{PMPSGLVTC22} (symbols joined by lines) for systems A--C2 at $T^*=1.5$. For better clarity,  the data have been successively shifted  upward  one unit.}
 \end{figure}

Before comparing with simulation results, let us first consider the low-density behavior. Figure \ref{fig:g1} displays the difference $g^{(1)}(r)-g^{(1)}_{\text{exact}}(r)$ at the reduced temperatures $T^*\equiv k_BT/|\epsilon_1|=1.5$ and $2$ for the systems of Table \ref{table:acronym}. We can observe that the deviations of the RFA values from the exact ones are rather small for the SW system A. In the SW+SS cases, the deviations  tend to increase as one moves from system B1 to system B4. i.e., as the height  and/or  the range of the energy barrier increase. A similar behavior is observed in the SW+SS+SW (systems C1 and C2), this time by increasing the depth of the second well. Those deviations are substantially reduced if temperature increases. Interestingly, while the RFA overestimates $g^{(1)}(r)$ for system A, it underestimates it for systems B1--B4. In the case of systems C1 and C2, the difference $g^{(1)}(r)-g^{(1)}_{\text{exact}}(r)$ changes from negative to positive at a certain distance intermediate between $\lambda_2$ and $\lambda_3$. Regarding the jumps $\Delta g^{(1)}(\lambda_j)$, we can observe that the RFA underestimates $\Delta g^{(1)}(1)$, yields the exact $\Delta g^{(1)}(\lambda_n)$, and overestimates $\Delta g^{(1)}(\lambda_j)$ for $1\leq j \leq n-1$ in systems B1--C2.

The deviations $b_{3,v}-b_{3,\text{exact}}$ and $b_{3,c}-b_{3,\text{exact}}$ are shown in Fig.~\ref{fig:b3} for the temperature range $1\leq T^*\leq 2$. As happened in Fig.~\ref{fig:g1}, the magnitudes of those deviations generally decrease with growing temperature and increase when going from system A to system B4 and from system C1 to system C2. We have noted that this is generally the case not only for the absolute deviations but also for  the relative ones. For instance, the magnitudes of the relative deviations of the pair $(b_{3,v}, b_{3,c})$ with respect $b_{3,\text{exact}}$ at $T^*=1.5$ are $(0.1\%,0.2\%)$, $(0.3\%,0.3\%)$, $(1.6\%,1.9\%)$, $(2.7\%,4.4\%)$, $(6.6\%,9.1\%)$, $(2.7\%,5.8\%)$, and $(1.6\%,17.7\%)$ for systems A--C2, respectively. For system A, both $b_{3,v}$ and $b_{3,c}$ underestimate the exact values, while the opposite occurs for systems B1--B4. In the case of systems C1 and C2, $b_{3,v}$ overestimates $b_{3,\text{exact}}$ but $b_{3,c}$ underestimates it. The virial-route value $b_{3,v}$ is more accurate than the compressibility-route value $b_{3,c}$ in systems A--B4 and C2. In contrast, in the case of system C1, $b_{3,c}$ becomes more accurate than $b_{3,v}$ if $T^*>1.24$.

Now we turn to the comparison with the recent simulation data of Perdomo {et al.} \cite{PMPSGLVTC22}, starting with the structural properties. In Fig.~\ref{fig:g_A-B}  we show the comparison between the results of the RFA approach for the RDF and the  simulation values for systems A--B4 at different conditions of  reduced density $\rho^*$ and reduced temperature $T^*$. There is clearly a very good agreement between the theoretical and the simulation results in all cases. On the other hand, Fig.~\ref{fig:g_A-B}(a) shows that in the chosen states of systems B3 and B4, the RFA does not capture some subtle details of the RDF in the interval $\lambda_2=1.5\leq r\leq 2$.

The general good agreement found in Fig.~\ref{fig:g_A-B} is noteworthy for two reasons. First, taking into account that the values of the density are not small ($\rho^*=0.6$ and $0.9$, corresponding to $\eta=0.31$ and $0.47$, respectively), the agreement is clearly better than one might have expected in view of Figs.~\ref{fig:g1}(a--e). Second, as said in Section~\ref{sec1}, the simulation results do not correspond to genuine discontinuous potentials but to continuous potentials that mimic the discontinuous ones (see Ref.~\cite{PMPSGLVTC22} for details). The latter feature manifests itself in the fact that, in contrast to the RFA results, the RDF simulation data do not present jumps at $r=\lambda_j$, as clearly seen in Fig.~\ref{fig:g_A-B}(b).

A more stringent test is carried out in Fig.~\ref{fig:g_C}, where the three-step system C2 is considered for the  temperature $T^*=1$. A generally good agreement, comparable to the one already observed in Fig.~\ref{fig:g_A-B}, is present up to $\rho^*=0.6$. However, the agreement tends to deteriorate for the highest densities.

Within the RFA scheme, it is straightforward to obtain the static structure factor $S(q)$ through Eq.~\eqref{b1}. In Fig.\ \ref{fig:S_k}, we illustrate the behavior of  $S(q)$ as a function of the wavenumber (divided by $2\pi$) at the state point $\rho^*=0.6$, $T^*=0.7$ for all cases listed in Table \ref{table:acronym}. As expected, $S(k)$ presents peaks at values of $q/2\pi$ around integer numbers, thus signaling the order induced by the hard-core diameter. In the particular case of system B4 (which corresponds to  the highest and widest repulsive barrier), an ``anomalous'' local peak  in the structure factor at $q/2\pi \sim 1/\lambda_2$ is clearly present. The existence of such a peak  associated with competing interactions has been previously described \cite{GCC07}. In fact, in the case of SW+SS systems, the limits $T^*\to\infty$ and $T^*\to 0$ correspond to hard-sphere fluids of diameters $\sigma=1$ and $\lambda_2$, respectively. Thus,  the local peak of $S(q)$ at $q/2\pi \sim 1/\lambda_2$ in system B4 becomes less pronounced as temperature increases and eventually disappears at about $T^*\simeq 0.9$ if $\rho=0.6$ (not shown).

After having analyzed the structural properties, we present in Fig.~\ref{fig:Z} the comparison between the results for the compressibility factor as a function of the reduced density $\rho^*$ computed using the RFA (taking both the virial and the compressibility routes) and the corresponding simulation data \cite{PMPSGLVTC22} at  a reduced temperature $T^*=1.5$.
One can observe that both RFA predictions agree well with the simulation data up to $\rho^*=0.6$ (systems A--B3) or $\rho^*=0.4$ (systems B4--C2). This represents a performance of the RFA better than expected from the behavior of the third virial coefficient in Fig.~\ref{fig:b3}.
In the high-density region ($\rho^*>0.6$) we see that both the RFA virial and compressibility routes tend to underestimate $Z$ in the case of the systems A--B4, what contrasts with the behavior observed in Fig.~\ref{fig:b3} for the third virial coefficient of systems B1--B4. On the other hand, in the case of the three-step systems C1 and C2, the virial (compressibility) route overestimates (underestimates) the compressibility factor, this time in qualitative agreement with Figs.~\ref{fig:b3}(f,g). It is also quite apparent from Fig.~\ref{fig:Z} that the RFA compressibility route is more reliable than the RFA virial route for systems A--B4, but the opposite occurs for systems C1 and C2.

It should  also be noted from Fig.~\ref{fig:Z} that, in all cases, when the values of the compressibility factor $Z$ obtained by the compressibility and virial RFA routes  agree with each other, then they agree with the simulation results. This then provides a useful criterion to estimate  the compressibility factor from the RFA method if simulation data are absent.

\section{Concluding remarks}
\label{concl}
In  this paper we have  used the RFA approach introduced earlier for a general potential with a hard core and $n$-step constant sections \cite{SYH12} to compute the structural properties and the equation of state of a particular kind of discrete-potential fluids, namely the ones in which molecules interact via a potential consisting of a hard core plus up to  three constant sections of different heights and widths, the middle one being a square shoulder. In the end,  the method requires the (numerical) solution of  $n=1$ (system A), $n=2$ (systems B1--B4), or $n=3$ (systems C1 and C2) coupled transcendental equations. In any case, being a semi-analytical nonperturbative method, the RFA naturally presents a clear advantage over other approaches, such as the usual integral equation method in the theory of liquids.

As a supplement to the results first derived in Ref.~\cite{SYH12}, we have obtained in this paper the RFA and exact analytical expressions of the RDF to first order in density and of the third virial coefficient for the general $n$-step interaction potential. The use of the RFA to find the equation of state of piecewise constant potentials is another contribution of the present work.

Apart from the comments made in the previous section concerning the cases we chose to illustrate our results, some additional remarks are in order.
To begin with, the comparison with the most recent simulation data reinforces the conclusions drawn in our previous papers \cite{SYH12,SYHBO13}, namely that the RFA approach is a valuable tool to compute the structural properties of fluids whose molecules interact via potentials consisting of a hard-core followed by one or more constant sections, except perhaps at low temperatures and high densities. In the present paper we have also shown that the compressibility factor of such systems derived with the RFA, using either the virial or the compressibility routes, is also reasonably accurate. While in principle one might try to obtain with the present approach also other thermodynamic properties, such as the liquid-vapor coexistence of these fluids, the heavy numerical work involved places this task outside the scope of this  paper. Finally, the good agreement between our results for the RDF and the simulation data (obtained with an equivalent continuous potential) \cite{PMPSGLVTC22} suggests that, at least for the structural and thermodynamic properties, replacing discrete potentials with (appropriate) continuous ones \cite{MCSM22} does not lead to serious errors.



\emph{CRediT authorship contribution statement}

% See https://www.elsevier.com/authors/policies-and-guidelines/credit-author-statement

\textbf{S. B. Yuste}: Conceptualization, Methodology, Software, Formal analysis, Writing - Review \& Editing, Funding acquisition
\textbf{A. Santos}: Conceptualization, Methodology, Formal analysis, Writing - Review \& Editing, Visualization, Funding acquisition
\textbf{M. L\'opez de Haro}: Conceptualization, Methodology, Writing - Original Draft

\section*{Declaration of Competing Interest}

The authors declare that they have no known competing financial
interests or personal relationships that could have appeared
to influence the work reported in this paper.

\section*{Acknowledgments}
We are grateful to Dr.~Ram\'on Casta{\~n}eda-Priego for providing us with the simulation data employed in Figs.~\ref{fig:g_A-B}--\ref{fig:Z}.
S.B.Y. and A.S.\ acknowledge financial support from Grant PID2020-112936GB-I00 funded by MCIN/AEI/10.13039/501100011033, and from Grants IB20079 and GR21014 funded by Junta de Extremadura (Spain) and by ERDF ``A way of making Europe.''

 %\bibliographystyle{elsarticle-num}


%\section*{References}

\bibliography{C:/AA_D/Dropbox/Mis_Dropcumentos/bib_files/liquid}
\end{document}